\documentclass[11pt]{article}
\usepackage{jmlr2e}
\usepackage{times}
\usepackage{amsmath}
\usepackage{amssymb}
\usepackage{color}
\usepackage{graphicx}
\usepackage{natbib}
\usepackage{algorithmic}
\usepackage{algorithm}
\usepackage{setspace}
\usepackage{bbm}

\newcommand{\R}{\mathbbm{R}}

\DeclareMathOperator*{\argmin}{arg\,min}
\newcommand{\half}{\textstyle{\frac{1}{2}}}

\firstpageno{1}

\begin{document}

\title{Alternating Linearization for Structured Regularization Problems}

\author{\name Xiaodong Lin \email lin@business.rutgers.edu \\
       \addr Department of Management Science and Information Systems\\
       Rutgers University\\
       Piscataway, NJ 08854
       \AND
       \name Minh Pham \email ptuanminh@gmail.com \\
       \addr Statistical and Applied Mathematical Sciences Institute (SAMSI)\\
       Durham, NC 27707
       \AND
       \name Andrzej Ruszczy\'nski \email rusz@business.rutgers.edu \\
       \addr Department of Management Science and Information Systems\\
       Rutgers University\\
       Piscataway, NJ 08854
       }

%\editor{XXX}

\maketitle

%\today
%\maketitle
%\thankstext{T1}{Footnote to the title with the `thankstext' command.}

\begin{abstract}
We adapt the alternating linearization method for proximal decomposition to structured regularization problems, in particular, to the generalized lasso
problems. The method is related to two well-known operator splitting methods, the Douglas--Rachford and the Peaceman--Rachford method, but it has descent
properties with respect to the objective function. This is achieved by employing a special update test, which decides
whether it is beneficial to make a Peaceman--Rachford step, any of the two possible
Douglas--Rachford steps, or none.
The convergence mechanism of the method is related to that of bundle methods of nonsmooth optimization.
We also discuss implementation for very large problems, with the use of specialized algorithms and sparse data structures.
Finally, we present numerical results for several synthetic and real-world examples, including a three-dimensional fused lasso problem, which
illustrate the scalability, efficacy, and accuracy of the method.
\end{abstract}


\section{Introduction}
\label{sec:intro}

Regularization techniques that encourage sparsity in parameter estimation have gained increasing popularity recently. The most widely used example is \emph{lasso} \citep{Tib96}, where the loss function $f(\cdot)$ is penalized by the $\ell_1$-norm of the unknown coefficients
$\beta\in \R^p$, to form a modified objective function,
\begin{equation}
\mathcal{L}(\beta) = f(\beta) + \lambda \|\beta \|_1, \quad \lambda >0,
\end{equation}
in order to shrink irrelevant coefficients to zero. Many efficient algorithms have been proposed to solve this problem, including \citep{Fu98, Daub04, Efron04} and \citep{Friedman07}. Some of them are capable of handling massive data sets with tens of thousands of variables and observations.

For many practical applications, physical constraints and domain knowledge may mandate additional structural constraints on the parameters. For example, in cancer research, it may be important to consider groups of interacting genes in each pathway rather than individual genes. In image analysis, it is natural to regulate the differences between neighboring pixels in order to achieve smoothness and reduce noise. In light of these popular demands, a variety of structured penalties have been proposed to incorporate prior information regarding model parameters. %\citep{Yuan06} proposed to penalize pre-specified non-overlapping groups of coefficients, so that one can shrink all the coefficients within the group to zero. \citep{Kim10} extended this to overlapping cases based on a tree structure.
One of the most important structural penalties is the \emph{fused lasso} proposed in \citep{Tib05}. It utilizes the natural ordering of input variables to achieve parsimonious parameter estimation on neighboring coefficients. \cite{Chen10}  developed the \emph{graph induced fused lasso} that penalizes differences between coefficients associated with nodes in a graph that are connected. \cite{Beck09} proposed the \emph{total variation penalty} for image denoising and deblurring, in a similar fashion to the two-dimensional fused lasso. Similar penalty functions have been successfully applied to several neuroimaging studies \citep{Michel11, Grosenick11, Taylor13}. More recently, \cite{Zhang12} applied a generalized version of fused lasso to reconstruct gene copy number variant regions. %\cite{Taylor13} proposed a variants of graph-constrained Elastic Net models that impose structural constrained on the model coefficients for whole-brain fMRI analysis.
A general structural lasso framework was proposed in \citep{tib11}, with the following form:
 \begin{equation}
  \label{glasso}
\mathcal{L}(\beta) = f(\beta) + \lambda \|R \beta \|_1, \quad \lambda >0,
\end{equation}
where $R$ is an $m \times p$ matrix that defines the structural constraints one wants to impose on the coefficients. Many regularization problems, including high dimensional fused lasso and graph induced fused lasso, can be cast in this framework.

When the structural matrix $R$ is relatively simple, as in the original lasso case with $R=I$, traditional path algorithms and coordinate descent techniques can be used to solve the optimization problems efficiently  \citep{Friedman07}. For more complex structural regularization,  these methods cannot be directly applied. One of the key difficulties is the non-separability of the nonsmooth penalty function. Coordinate descent methods fail to converge under this circumstances \citep{Tseng01}. Generic solvers, such as interior point  methods, can sometimes be used; unfortunately they become increasingly inefficient for large size problems, particularly when the design matrix is ill-conditioned \citep{Chen11}.

In the past two years, many efforts have been devoted to developing efficient optimization techniques for solving regularization problems using structured penalties. \cite{Jun10} and \cite{Ye-Xie} developed a first-order and a split Bregman scheme, respectively, for solving similar class of problems.
 \cite{Chen11} proposed a modified proximal technique for the general structurally penalized problems. It is based on a first order approximation of the nonsmooth penalty function, which can become unstable when dimension is high. Meanwhile, several path algorithms have also been proposed to compute the whole regularization path for the general fused lasso problem.  \cite{Hoefling10} developed a path algorithm for solving (\ref{glasso}) when the matrix $X^T X$ is nonsingular. This technique is not applicable to cases with large dimension of $\beta$ and small number of observations, such as gene expression and brain imaging analysis. \cite{tib11} extended the path algorithm to include all design matrices $X$, by computing the regularization path of the dual problem. Although  fairly general, this version of the path algorithm does not scale well with data dimension, as the knots of the piecewise linear solution path become very dense.
Many of the proposed approaches are versions of the \emph{operator splitting methods} or their dual versions, \emph{alternating direction methods}
(see, e.g., \cite{Boyd-et-al,Combettes-Pesquet}, and the references therein). Although fairly general and universal, they frequently suffer from slow tail convergence
(see \citep{He-Yuan} and the references therein).

 Thus, a need arises to develop a general approach that can solve large scale structured regularization problem efficiently. For such an approach to be successful in practice, it should guarantee to converge at a fast rate, be able to handle massive data sets, and should not rely on approximating the penalty function.
 In this paper, we propose a framework based on the alternating linearization algorithm of \citep{Kiwiel99}, that satisfies all these requirements.

  We consider the following generalization of \eqref{glasso}:
  \begin{equation}
  \label{glasso1}
\mathcal{L}(\beta) = f(\beta) + \lambda \|R \beta \|_\lozenge, \quad \lambda >0,
\end{equation}
where $\|\cdot\|_\lozenge$ is a norm in $\R^m$. Our considerations and techniques will apply to several possible
choices of this norm, in particular, to the $\ell_1$ norm $\|\cdot\|_1$, and to the total variation norm $\|\cdot\|_{\text{TV}}$ used in image processing.

Formally, we write the objective function as a sum of two convex functions,
\begin{equation}\label{la}
\mathcal{L}(\beta)= f(\beta) + h(\beta),
\end{equation}
where $f(\beta)$ is a loss function, which is assumed to be convex with respect to $\beta$,  and $h(\cdot)$ is a convex penalty function. Any of the functions (or both) may be nonsmooth, but an essential requirement of our framework is that each of them  can be easily minimized with respect to $\beta$, when augmented by a separable linear-quadratic term $\sum_{i=1}^p \left(s_i\beta_i + d_i\beta_i^2\right)$, with some vectors $s,d\in \R^p$, $d >0$.
Our method bears resemblance  to operator splitting and
alternating direction approaches, but differs from them in the fact that it is \emph{monotonic} with respect to the values of \eqref{la}. We discuss these
relations and differences later in section \ref{sec:relations}, but roughly speaking, a special test applied at every iteration of the method decides which of
the operator splitting iterations is the most beneficial one.

In our applications, we focus on the quadratic loss function $f(\cdot)$ and the penalty function in the form of generalized lasso (\ref{glasso1}), as the most important case, where comparison with other approaches is available.  This case satisfies the requirement specified above, and allows for substantial specialization and acceleration of the general framework of alternating linearization.
In fact, it will be clear from our presentation that any convex loss function $f(\cdot)$ can be handled in exactly the same way.

An important feature of our approach is that problems with the identity design matrix are solved exactly in one iteration, even for very large dimension.

The remainder of the paper is organized as follows. %In Section 2, we describe a few examples of the generalized lasso problem.
In Section \ref{sec:AL}, we introduce the alternating linearization method and we discuss its relations to other approaches.
Section \ref{sec:lasso} briefly discusses the application to lasso problems. In section \ref{sec:gen-lasso} we describe the application to
general structured regularization problems.
Section \ref{s:numerical} presents sim\-ulation results and real data examples, which illustrate the  efficacy, accuracy, and scalability of the
alternating linearization method.  Concluding remarks are presented in section \ref{s:conclusions}. The appendix contains details about the
algorithms used to solve the subproblems of the alternating linearization method.

\section{The alternating linearization method}
\label{sec:AL}

\subsection{Outline of the method}

In this section, we describe the alternating linearization (ALIN) approach to minimize (\ref{la}).
It is an iterative method, which generates a sequence of approximations $\{\hat{\beta}^k\}$ converging to a solution of the
original problem~(\ref{la}), and two auxiliary sequences: $\{\tilde{\beta}_h^k\}$ and $\{\tilde{\beta}_f^k\}$, where $k$ is the iteration number.
Each iteration of the ALIN algorithm consists of solving two subproblems: the \emph{$h$-subproblem} and the \emph{$f$-subproblem}, and of an \emph{update step}, applied after any of the subproblems, or after each of them.

At the beginning we set $\tilde{\beta}_f^0=\hat{\beta}^0$, where $\hat{\beta}^0$ is the starting point of the method. In the description below, we suppress the superscript $k$ denoting the iteration number, to simplify notation.

\subsubsection*{\emph{The $h$-subproblem}}
We linearize $f(\cdot)$ at $\tilde{\beta}_f$, and approximate it by the function
\[
\tilde{f}(\beta) = f(\tilde{\beta}_f) + s_f^T(\beta - \tilde{\beta}_f).
\]
If $f(\cdot)$ is differentiable, then $s_f =\nabla f(\tilde{\beta}_f)$; for a general convex $f(\cdot)$, we select a subgradient $s_f\in \partial f(\tilde{\beta}_f)$.
In the first iteration, this may be an arbitrary subgradient; at later iterations special selection rules apply, as described in \eqref{sf} below.

The approximation is used in the optimization problem
\begin{equation}
\label{hproblem}
\min_{\beta}\  %\mathcal{L}_h(\beta) =
    \tilde{f}(\beta) + h(\beta) + {\half} \| \beta - \hat{\beta} \|_{D}^2,
\end{equation}
in which the last term is defined as follows:
\[
   \| \beta - \hat{\beta} \|_{D}^2 = (\beta - \hat{\beta})^T D (\beta - \hat{\beta}),
\]
with a diagonal matrix $D=\textup{diag}\{d_j,\, j=1,\dots,p\}$, $d_j>0$, $j=1,\dots,p$.
The solution of the $h$-sub\-problem (\ref{hproblem}) is denoted by $\tilde{\beta}_h$.

We complete this stage by calculating the subgradient of $h(\cdot)$ at $\tilde{\beta}_h$, which features in the
optimality condition for the minimum in (\ref{hproblem}):
\[
0 \in  s_f + \partial h(\tilde{\beta}_h) + D(\tilde{\beta}_h-\hat{\beta}).
\]
Elementary calculation yields the right subgradient $s_h\in \partial h(\tilde{\beta}_h)$:
\begin{equation}
\label{sh}
{s}_h = -{s}_f-D (\tilde{\beta}_h-\hat{\beta}).
\end{equation}
\subsubsection*{\emph{The $f$-subproblem}}
Using the subgradient $s_h$ we construct a linear minorant of the penalty function $h(\cdot)$ as follows:
\[
\tilde{h}(\beta) = h(\tilde{\beta}_h) + s_h^T(\beta - \tilde{\beta}_h).
\]
This approximation is employed in the optimization problem
\begin{equation}
\label{fproblem}
 \min_{\beta} \ % \mathcal{L}_f (\beta) =
 f(\beta) +\tilde{h}(\beta) + {\half} \| \beta - \hat{\beta} \|_{D}^2.
 \end{equation}
The optimal solution of this problem is denoted by $\tilde{\beta}_f$. It will be used in the next iteration as the
point at which the new linearization of $f(\cdot)$ will be constructed. The next subgradient of $f(\cdot)$ to be used in the
$h$-subproblem will be
\begin{equation}
\label{sf}
{s}_f = -{s}_h-D (\tilde{\beta}_f-\hat{\beta}).
\end{equation}

\subsubsection*{\emph{The update step}}
The update step can be applied after any of the subproblems, or after both of them. It changes the current best
approximation of the solution $\hat{\beta}$, if certain improvement conditions are satisfied. It uses a parameter $\gamma\in (0,1)$. In the implementation, we use $\gamma=0.2$. The choice of this parameter does not influence the overall performance of the algorithm.
We describe it here for the case of applying the update step after the $f$-subproblem; analogous operations are
carried out if the update step is applied after the $h$-subproblem.

At the beginning of the update step the stopping criterion is verified. If
\begin{equation}
\label{stopping}
f(\tilde{\beta}_f) + \tilde{h}(\tilde{\beta}_f) \ge f(\hat{\beta}) + h(\hat{\beta}) - \varepsilon,
\end{equation}
the algorithm terminates. Here $\varepsilon>0$ is the stopping test parameter.

If the the stopping test is not satisfied, we check the inequality
\begin{equation}
\label{update}
f(\tilde{\beta}_f) + h(\tilde{\beta}_f) \le (1-\gamma) \big[f(\hat{\beta}) + h(\hat{\beta})\big]
+ \gamma\big[ f(\tilde{\beta}_f) + \tilde{h}(\tilde{\beta}_f)\big].
\end{equation}
If it is satisfied,
then we update $\hat{\beta}\gets \tilde{\beta}_f$; otherwise $\hat{\beta}$ remains unchanged.

If the update step is applied after the $h$-subproblem, we use $\tilde{\beta}_h$ instead if $\tilde{\beta}_f$ in the inequalities \eqref{stopping}
and~\eqref{update}.

The update step is a crucial component of the alternating linearization algorithm; it guarantees that the sequence
$\{\mathcal{L}(\hat{\beta}^k)\}$ is monotonic, and it stabilizes the entire algorithm (see the remarks at the end of sect\-ion~\ref{s:simulation}). It is a specialized form of the main distinction between \emph{null} and \emph{serious} steps
in \emph{bundle methods} for nonsmooth optimization. The Reader may consult the  books \cite{bogilesa:on,HUL-book,Kiwiel-book,Rusz-book}
and the references therein for the theory of bundle methods and the significance of null and serious steps in these methods.

\subsection{Relation to operator splitting and alternating direction methods}
\label{sec:relations}

Our approach is intimately related to \emph{operator splitting methods} and their dual versions, \emph{alternating direction methods},
which are recently very popular in the area of signal processing (see, e.g., \citep{Boyd-et-al,Combettes-Pesquet,Fadili-Peyre}).
To discuss these relations, it is convenient to present our method formally, and to introduce
two running \emph{proximal centers}:
\begin{align*}
z_f = \hat{\beta} -D^{-1}s_f,\\
z_h = \hat{\beta} -D^{-1}s_h.
\end{align*}
After elementary manipulations we can absorb the linear terms into the quadratic terms and summarize the alternating linearization method as follows.
\begin{algorithm}[H]
    \caption{Alternating Linearization}
    \label{AL-alg}
    \begin{algorithmic}[1]
    \REPEAT
        \STATE $\tilde{\beta}_h \leftarrow \argmin\big\{ h(\beta) + \frac{1}{2} \| \beta - z_f \|^2_D\big\}$
        \IF{(\textit{Update Test for} $\tilde{\beta}_h$)} \label{test1}
            \STATE $\hat{\beta}\gets \tilde{\beta}_h$ \label{update1}
        \ENDIF
        \STATE $z_h \gets \hat{\beta} + \tilde{\beta}_h - z_f$
       \STATE $\tilde{\beta}_f \leftarrow \argmin\big\{ f(\beta) + \frac{1}{2} \| \beta - z_h \|^2_D\big\}$
       \IF{(\textit{Update Test for} $\tilde{\beta}_f$)} \label{test2}
            \STATE $\hat{\beta}\gets \tilde{\beta}_f$ \label{update2}
        \ENDIF
        \STATE $z_f \gets \hat{\beta} + \tilde{\beta}_f - z_h$
        \UNTIL{(\textit{Stopping Test})}
    \end{algorithmic}
\end{algorithm}
The \emph{Update Test} in lines \ref{test1} and \ref{test2} is the corresponding version of inequality \eqref{update}.
The \emph{Stopping Test} is inequality \eqref{stopping}.

If we assume that the update steps in lines \ref{update1} and \ref{update2}  are carried out after \emph{every} $h$-subproblem
and \emph{every} $f$-subproblem, {without} verifying the update test \eqref{update}, then the method becomes equivalent to a scaled version of the
Peaceman--Rachford algorithm (originally proposed by \citep{Peaceman-Rachford} for PDEs and later generalized and analyzed by \citep{Lions-Mercier};
see also  \citep{Combettes2009} and the references therein). If $D=\rho I$ with $\rho>0$, then we obtain an unscaled version of this algorithm.

If we assume that the update steps are carried out after \emph{every} $h$-subproblem
{without} verifying inequali\-ty~\eqref{update}, but \emph{never} after $f$-subproblems, then the method becomes equivalent to a scaled version of the
Douglas--Rachford algorithm (introduced by \citep{Douglas-Rachford}, and generalized and analyzed by \citep{Lions-Mercier};
see also \citep{Bauschke-Combettes} and the references therein).
As the roles of $f$ and $h$ can be switched, the method in which updates are carried
always after $f$-subproblems, but never after $h$-subproblems, is also equivalent to a scaled Douglas--Rachford method.

Operator splitting methods are not monotonic with respect to the values of the objective function $\mathcal{L}(\beta)$. Their convergence is based
on monotonicity with respect to the distance to the optimal solution of the problem \citep{Lions-Mercier,Eckstein-Bertsekas}.

 In contrast, the convergence
mechanism of our method is different; it draws from some ideas of bundle methods in nonsmooth optimization
 \citep{HUL-book,Kiwiel-book,Rusz-book}.
Its key element is the
update test employed in \eqref{update}. At every iteration we decide whether it is beneficial to make a Peaceman--Rachford step, any of the two possible
Douglas--Rachford steps, or none. In the latter case, which we call the \emph{null step}, $\hat{\beta}$ remains unchanged, but the trial points
$\tilde{\beta}_h$ and $\tilde{\beta}_f$ are updated.
These updates continue, until $\tilde{\beta}_h$ or $\tilde{\beta}_f$ become better than $\hat{\beta}$, or until optimality is detected
(\emph{cf}. the remarks at the end of sect\-ion~\ref{s:simulation}).
In may be worth noticing that the recent application of the idea of alternating linearization in \cite{GMS-2012} removes the
update test from the method of \cite{Kiwiel99}, thus effectively reducing it to an operator splitting method.

%This additionally stabilizes the iterates, accelerates convergence, and improves accuracy.

\emph{Alternating direction methods} are dual versions
of the operator splitting methods, applied to the following equivalent form of the problem of minimizing \eqref{la}:
\begin{equation}\label{la-dual}
\min  f(\beta_1) + h(\beta_2),\quad \text{subject to} \quad \beta_1=\beta_2.
\end{equation}
In regularized signal processing problems, when $f(\beta) = \varphi(X\beta)$ with some fixed matrix $X$, the convenient problem formulation is
\[
\min  \varphi(v) + h(\beta),\quad \text{subject to} \quad v=X \beta.
\]
The dual functional,
\[
L_D(\lambda) = \min_{v} \big\{ \varphi(v) - \lambda^T v \big\}  + \min_{\beta} \big\{ h(\beta) + \lambda^TX\beta \big\},
\]
has the form of a sum of two functions, and the operator splitting methods apply.
The reader may consult \citep{Boyd-et-al,Combettes-Pesquet} for appropriate derivations.
It is also worth mentioning that the alternating direction methods are sometimes called
\emph{split Bregman methods} in the signal processing literature
(see, e.g., \citep{Goldstein-Osher,Ye-Xie}, and the references therein). Recently, \citep{QinGoldfarb} applied alternating direction methods
to some structured regularization problems resulting from group lasso models.

However, to apply our alternating linearization method to the dual problem of maximizing $L_D(\lambda)$, we would have to be able to quickly compute
the value of the dual functions, in order to verify the update condition \eqref{update}, as discussed in detail in \cite{Kiwiel99}. The second dual
function,  $\min_{\beta} \big\{ h(\beta) +\lambda^T X\beta \big\}$ is rather difficult to evaluate, and it makes the update test time consuming. Without this test,
our method reduces to the alternating direction method, which does not have descent properties, and whose tail convergence may be slow.
Our experiments reported at the end of section \ref{s:simulation} confirm these observations.

\subsection{Convergence}

Convergence properties of the alternating linearization method follow from the general theory developed in \citep{Kiwiel99}. Indeed, after
the change of variables $\xi = D^{1/2}\beta$ we see that the method is identical to Algorithm 3.1 of \citep{Kiwiel99},
with $\rho_k = 1$. The following statement is a direct consequence of \citep[Theorem 4.8]{Kiwiel99}.
\begin{theorem}
\label{t:1}
Suppose that the set of minima of the function {\rm (\ref{la})} is nonempty. Then the sequence $\{\hat{\beta}^k\}$ generated by the
algorithm is convergent to a minimum point $\beta^*$ of the function  {\rm (\ref{la})}. Moreover, every accumulation point
$(s_f^*,s_h^*) $ of the
sequence $\{ (s_f^k,s_h^k) \}$ satisfies the relations: $s_f^*\in \partial f(\beta^*)$, $s_h^*\in \partial h(\beta^*)$, and
$s_f^*+s_h^*=0$.
\end{theorem}

For structured regularization problems the assumption of the theorem is satisfied, because both the loss function $f(\cdot)$
and the regularizing function $h(\cdot)$ are bounded from below, and one of the purposes of the regularization term is to make the
set of minima of the function $\mathcal{L}(\cdot)$ nonempty and bounded.

It may be of interest to look closer at the stopping test \eqref{stopping} employed in the update step.

\begin{lemma}
\label{l:stopping}
Suppose $\beta^*$ is the unique minimum point of $\mathcal{L}(\cdot) = f(\cdot) + h(\cdot)$ and let  $\alpha>0$ be such that
$\mathcal{L}(\beta) - \mathcal{L}(\beta^*) \ge \alpha \|\beta - \beta^*\|_D^2$ for all $\beta$. Then the stopping criterion \eqref{stopping}
implies that
\begin{equation}
\label{val-est}
\mathcal{L}(\hat{\beta}) - \mathcal{L}(\beta^*) \le \frac{\varepsilon}{\alpha}.
\end{equation}
\end{lemma}
\begin{proof}
 As $\tilde{h}(\cdot) \le h(\cdot)$, inequality \eqref{stopping} implies that
\begin{equation}
\label{stopping2}
\begin{aligned}
\min_{\beta} \Big\{ f(\beta) + h(\beta) + \frac{1}{2}\|\beta - \hat{\beta}\|_D^2 \Big\} &\ge
\min_{\beta} \Big\{ f(\beta) + \tilde{h}(\beta) + \frac{1}{2}\|\beta - \hat{\beta}\|_D^2 \Big\}  \\
&=
f(\tilde{\beta}_f) + \tilde{h}(\tilde{\beta}_f) + \frac{1}{2}\|\tilde{\beta_f} - \hat{\beta}\|_D^2 \\
&\ge
f(\hat{\beta}) + h(\hat{\beta})   - \varepsilon.
\end{aligned}
\end{equation}
The expression on the left hand side of this inequality is the Moreau--Yosida regularization of the function
$\mathcal{L}(\cdot)$ evaluated at $\hat{\beta}$. By virtue of \cite[Lemma 7.12]{Rusz-book}, after setting $\tilde{x}=\beta^*$ and
with the norm $\|\cdot\|_D$, the Moreau--Yosida regularization satisfies the following inequality:
\[
\min_{\beta} \Big\{ \mathcal{L}(\beta) + \frac{1}{2}\|\beta - \hat{\beta}\|_D^2 \Big\} \le
\mathcal{L}(\hat{\beta})  - \frac{\big(\mathcal{L}(\hat{\beta}) - \mathcal{L}(\beta^*)\big)^2}{\| \hat{\beta} - \beta^*\|^2_D}.
\]
Combining this inequality with \eqref{stopping2} and simplifying, we conclude that
\[
\frac{\big(\mathcal{L}(\hat{\beta}) - \mathcal{L}(\beta^*)\big)^2}{\| \hat{\beta} - \beta^*\|^2_D} \le \varepsilon.
\]
Substitution of the denominator by the upper estimate $\big(\mathcal{L}(\beta) - \mathcal{L}(\beta^*)\big)/\alpha$ yields \eqref{val-est}.
\end{proof}

If $\beta^*$ is unique then $\mathcal{L}(\cdot)$ grows at least quadratically in the neighborhood of $\beta^*$. This implies that
 $\alpha>0$ satisfying the assumptions of Lemma \ref{l:stopping} exists.
 Our use of the norm $\|\cdot\|_D$ amounts to comparing the function $(\beta-\beta^*)^T X^TX(\beta-\beta^*)$ to
its diagonal approximation $(\beta-\beta^*)^T D (\beta-\beta^*)$ for $D=\text{diag}(X^TX)$. The reader may also consult \cite[Lemma 1]{Ruszczynski95} for the
accuracy of the diagonal approximation when the matrix $X$ is sparse.

If the original problem is to minimize \eqref{la} subject to the constraint
that $\beta\in B$ for some convex closed set $B$, we can formally add the indicator function of this set,
\[
\delta(\beta) = \begin{cases}
0&\text{if $\beta\in B$},\\
+\infty& \text{if $\beta\not\in B$},
\end{cases}
\]
to $f(\cdot)$ or to $h(\cdot)$ (whichever is more convenient). This will result in including the constraint in one of the
subproblems, and changing the subgradients accordingly. The theory of Kiwiel et al. (1999) covers this case as well,
and Theorem \ref{t:1} remains valid.

\section{Application to lasso regression}
\label{sec:lasso}

First, we demonstrate the alternating linearization algorithm (ALIN) on the classical lasso regression problem. Due to the separable nature
of the penalty function, very efficient coordinate descent methods are applicable to this problem as well \citep{Tseng01}, but we
wish to illustrate our approach on the simplest case first.

In the lasso regression problem we have
\[
f(\beta) = {\half} \|y-X\beta\|_2^2,\qquad h(\beta) = \lambda\|\beta \|_1,
\]
where $X$ is the $n \times p$ design matrix, $y\in \R^n$  is the vector of response variables, $\beta\in \R^p$ is the vector of regression coefficients, and  $\lambda>0$ is a parameter of the model.

We found it essential to use $D=\textup{diag}(X^TX)$, that is, $d_j = X_j^T X_j$, $j=1,\dots,p$.
This choice is related to the \emph{diagonal quadratic approximation} of the function  $f(\beta)=\frac{1}{2} \|y-X\beta\|_2^2$, which was employed
(for similar objectives in the context of augmented Lagrangian minimization) by \cite{Ruszczynski95}.
Indeed, in the $h$-subproblem in the formula \eqref{LAlasso} below, the quadratic regularization term is a quadratic form built on the diagonal
of the Hessian of $f(\cdot)$.% It is a good approximation of the original problem, especially when the design matrix $X$ is sparse \citep{Ruszczynski95}.

\subsubsection*{\emph{The $h$-subproblem}}

The problem (\ref{hproblem}), after skipping constants, simplifies to the following form
\begin{equation} \label{LAlasso}
\min_{\beta}\
 {s}_f^T \beta + \lambda\|\beta \|_1 + {\half} \| \beta- \hat{\beta} \|^2_{D},
\end{equation}
with ${s}_f = X^T (X \tilde{\beta}_f  - y )$.
Writing
$\tau_j=\hat{\beta}- {\tilde{s}_{fj}}/{d_j}$,
 we obtain the following closed form solutions of (\ref{LAlasso}), which can be calculated component-wise:
\begin{equation}
\label{f-lasso}
\tilde{\beta}_{hj}=\text{sgn}(\tau_j) \max\Big(0, |\tau_j|-\frac{\lambda}{d_j}\Big),\quad j = 1,\dots,p.
\end{equation}
 The subgradient ${s}_h$ of $h(\cdot)$ at $\tilde{\beta}_{h}$ is
calculated by (\ref{sh}).

\subsubsection*{\emph{The $f$-subproblem}}

The problem (\ref{fproblem}), after skipping constants, simplifies to the unconstrained quadratic programming problem
\begin{equation}
\label{opt-lasso}
 \min_{\beta} \ {s}_h^T \beta  + {\half}\|y-X\beta\|_2^2 + {\half}\|\beta-\hat{\beta}\|_{D}^2.
\end{equation}
Its solution can be obtained by solving the following symmetric linear system in $\delta=\beta-\hat{\beta}$:
\begin{equation}
\label{lin-lasso}
(X^TX+D)\delta = X^T(y-X\hat{\beta}) -{s}_h .
\end{equation}
This system can be efficiently solved by the preconditioned conjugate gradient method (see, e.g., \citep{Golub-VanLoan}),
with the diagonal preconditioner $D = \textup{diag}(X^TX)$. Its application does not require the explicit form of the matrix $X^TX$;
only matrix-vector multiplications with $X$ and $X^T$ are employed, and they can be implemented with sparse data structures.

The numerical accuracy of this approach is due to the good condition index of the resulting matrix, as explained in the following lemma.
\begin{lemma}
\label{l:conditioning}
The application of the preconditioned conjugate gradient method with preconditioner $D$ to the system \eqref{lin-lasso} is equivalent
to the application of the conjugate gradient method to a system with a symmetric positive definite matrix $\bar{H}$ whose
condition index is at most $\sqrt{p}+1$.
\end{lemma}
\begin{proof}
By construction, the preconditioned conjugate gradient method with a preconditioner $D$ applied to a system with a matrix $H$
is the standard conjugate gradient method applied to a system with the matrix $\bar{H} = D^{-\frac{1}{2}}H D^{-\frac{1}{2}}$.
Substituting the matrix from \eqref{lin-lasso}, we obtain:
\[
\bar{H} = D^{-\frac{1}{2}}( X^TX + D ) D^{-\frac{1}{2}} = D^{-\frac{1}{2}} X^TX D^{-\frac{1}{2}} + I.
\]
Define $\bar{X} = X D^{-\frac{1}{2}}$. By the construction of $D$, all columns $\bar{x}^j$, $j=1,\dots,p$, of $\bar{X}$ have
Euclidean length 1.

The condition index of $\bar{H}$ is equal to
\[
\text{cond}(\bar{H}) = \frac{ \lambda_{\max}(\bar{X}^T \bar{X})+1}{\lambda_{\min}(\bar{X}^T \bar{X})+1}.
\]
 As the matrix $\bar{X}^T \bar{X}$ is positive semidefinite, $\lambda_{\min}(\bar{X}^T \bar{X}) \ge 0$. To estimate
 $ \lambda_{\max}(\bar{X}^T \bar{X})$ suppose $v$ is the corresponding eigenvector of Euclidean length 1. We obtain the
 chain of relations:
 \[
 \sqrt{\lambda_{\max}(\bar{X}^T \bar{X})}  = \| \bar{X} v \|_2 = \Big\| \sum_{j=1}^p v_j \bar{x}^j \Big\|_2
 \le \sum_{j=1}^p |v_j| \| \bar{x}^j \|_2 = \|v\|_1 \le \sqrt{p}\|v\|_2 = \sqrt{p}.
 \]
Therefore, the condition index of $\bar{H}$ is at most $\sqrt{p}+1$.
\end{proof}

\section{Application to general structured regularization problems}
\label{sec:gen-lasso}

In the following we apply the alternating linearization algorithm to solve more general structured regularization problems including the generalized Lasso (\ref{glasso1}).
Here we assume the least square loss, as in the previous subsection. The objective function can be written as follows:
\begin{equation}
\label{GLR}
\mathcal{L}(\beta)=f(\beta) + h(\beta) ={\half} \|y-X\beta\|_2^2 + \lambda \| R \beta\|_\lozenge.
\end{equation}
For example, for the one-dimensional fused lasso, $R$ is the following $(p-1) \times p$ matrix:
\[
R=\begin{bmatrix}
-1 & 1 & 0 & \ldots & 0 \\
0 & -1 & 1 & \ldots & 0\\
%0 & 0 & -1 & 1 & \ldots \\
 \hdotsfor{5}\\
0 & 0 & \ldots & -1 & 1 \\
\end{bmatrix},
\]
and the norm $\|\cdot\|_\lozenge$ is the $\ell_1$-norm $\|\cdot\|_1$,
but our derivations are valid for any form of $R$, and any norm $\|\cdot\|_\lozenge$.

\subsubsection*{\emph{The $h$-subproblem}}

The $h$-subproblem can be equivalently formulated as follows:
 \begin{equation}
 \label{hs2}
\min_{\beta,z} \ {s}_f^T \beta + \lambda \|z\|_\lozenge +  {\half} \| \beta- \hat{\beta} \|^2_{D} \quad \text{subject to} \quad R\beta=z.
\end{equation}
Owing to the use of $D=\textup{diag}(X^TX)$, and with $s_f= X^T(X\hat{\beta}-y)$, it is a quite accurate approximation of the original problem,
especially for sparse $X$ \citep{Ruszczynski95}.

The Lagrangian of problem (\ref{hs2}) has the form
 \[
L(\beta,z,\mu)= {s}_f^T \beta +\lambda \|z\|_\lozenge +  \mu^T(R\beta-z)+{\half} \| \beta- \hat{\beta} \|^2_{D},
\]
where $\mu$ is the dual variable. Consider the dual norm $\|\cdot\|_*$, defined as follows:
\begin{equation}
\label{dual-norm}
\|\mu\|_* = \max_{\|z\|_\lozenge \le 1} \mu^T z,\qquad
\|z\|_\lozenge = \max_{\|\mu\|_* \le 1} \mu^T z.
\end{equation}
We see that
the minimum of the Lagrangian with respect to $z$ is finite if and only if
$\| \mu \| _{*} \leq  \lambda$  \cite[Example 2.94]{Rusz-book}.
 Under this condition, the minimum value of the $z$-terms is zero and we can eliminate them from the Lagrangian.
We arrive to its reduced form,
 \begin{equation}\label{fl-op}
\hat{L}(\beta,\mu)= {s}_f^T \beta +  \mu^T R\beta+{\half} \| \beta- \hat{\beta} \|^2_{D}.
\end{equation}
To calculate the dual function, we minimize $\hat{L}(\beta,\mu)$ over $\beta\in \R^p$. After elementary calculations, we obtain
the solution
\begin{equation}\label{fl-f}
\tilde{\beta}_h= \hat{\beta} - D^{-1}({s}_f+R^T\mu).
\end{equation}
Substituting it back to (\ref{fl-op}), we arrive to the following dual problem:
\begin{equation}
\label{Rdual}
\max_{\mu} \ -{\half}\mu^TRD^{-1}R^T\mu + \mu^T R(\hat{\beta}- D^{-1} {s}_f) \quad \text{subject to}\quad
\|\mu\|_{*} \leq \lambda.
\end{equation}
This is a norm-constrained optimization problem. Its objective function is quadratic, and the specific form
of the constraints depends on the norm $\|\cdot\|_\lozenge$ used in the regularizing term of \eqref{glasso1}.
\vspace{1ex}

\noindent
\emph{The case of the $\ell_1$-norm} \vspace{1ex}

If the norm $\|\cdot\|_\lozenge$ is the $\ell_1$-norm $\|\cdot\|_1$, then the dual norm is the $\ell_{\infty}$-norm:
\[
\|\mu\|_* = \|\mu\|_\infty = \max_{1\le j \le m} |\mu_j|.
\]
In this case \eqref{Rdual} becomes a box-constrained quadratic programming problem, for which
many efficient algorithms are available. One possibility is the
active-set box-constrained preconditioned conjugate gradient algorithm with spectral projected gradients, as described in \citep{Birgin-Martinez,Friedlander-Martinez}.
%The diagonal of the matrix $RD^{-1}R^T$ is a good preconditioner for this method
%in the applications that we dealt with.
%We summarize the most important
%operations of this method in the Appendix.
It should be stressed that its application does not require the explicit form of the matrix $RD^{-1}R^T$;
only matrix-vector multiplications with $R$ and $R^T$ are employed, and they can be implemented with sparse data structures.

An even better possibility, due to the separable form of the constraints,  is coordinate-wise optimization
(see, e.g., \cite[Sec. 5.8.2]{Rusz-book}) in the dual problem \eqref{Rdual}.
In our experiments, the dual coordinate-wise optimization
method strictly outperforms the box-constrained algorithm, in terms of the solution time.

The solution $\tilde{\mu}$ of the dual problem can be substituted into (\ref{fl-f}) to obtain the primal solution.
\vspace{1ex}

\noindent
\emph{The case of a sum of $\ell_2$-norms}\vspace{1ex}

Another important case arises when the vector $z=R\beta$ is split into $I$ subvectors $z^1,z^2,\dots,z^I$, and
\begin{equation}
\label{norm12}
\|z\|_\lozenge = \sum_{i=1}^I \|z^i\|_2.
\end{equation}
This is the group lasso model, also referred to as the $\ell_1/L_2$-norm regularization (see, e.g. \cite{QinGoldfarb}).
A special case of it is the \emph{total variation norm} is discussed in section \ref{s:TV}.

We can directly verify that the dual norm has the following form:
\[
\|\mu\|_* = \max_{1\le i \le I} \|\mu^i\|_2.
\]
It follows that problem \eqref{Rdual} is a block-quadratically constrained quadratic optimization problem:
\begin{equation}
\label{Rdual2}
\begin{aligned}
\max_{\mu} \ & -{\half}\mu^TRD^{-1}R^T\mu + \mu^T R(\hat{\beta}- D^{-1} {s}_f)\\
\text{s. t.}\ &
\|\mu^i\|_{2}^2 \leq \lambda^2, \quad i=1,\dots,I.
\end{aligned}
\end{equation}
This problem can be very efficiently solved by a cyclical block-wise optimization with respect to
the subvectors $\mu^1,\mu^2,\dots,\mu^I$. At each iteration of the method, optimization with respect to the corresponding
subvector $\mu^j$ is performed, subject to one constraint $\|\mu^j\|_{2}^2 \leq \lambda^2$. The other subvectors,
$\mu^i$, $i\ne j$ are kept fixed on their last values. After that, $j$ is incremented (if $j<I$) or reset to 1
(if $j=I$), and the iteration continues. The method stops when no significant improvements over $I$ steps can be observed. The dual block optimization method performs well in the applications
we are interested in. General convergence theory can be found in \citep{Tseng01}.

Again, the solution $\tilde{\mu}$ of the dual problem is substituted into (\ref{fl-f}) to obtain the primal solution.

\subsubsection*{\emph{The $f$-subproblem}}

We obtain the update $\tilde{\beta}_f$ by solving the linear equation system (\ref{lin-lasso}), exactly as in the lasso case.

\subsubsection*{\emph{The special case of $X=I$}}

If the design matrix $X=I$ in (\ref{GLR}), then our method solves the problem in one iteration,
when started from $\hat{\beta}=y$. Indeed, in this case we have $s_f=0$, $D=I$, and the first $h$-subproblem becomes equivalent to the original problem (\ref{GLR}):
 \begin{equation}
 \label{hs3}
\min_{\beta,z} \  \lambda \|z\|_\lozenge +  {\half} \| \beta- y \|_2^2 \quad \text{subject to} \quad R\beta=z.
\end{equation}
The dual problem (\ref{Rdual}) simplifies as follows:
\begin{equation}
\label{Rdual3}
\max_{\mu} \ -{\half}\mu^T RR^T\mu + \mu^T Ry \quad \text{subject to}\quad \|\mu\|_{*} \leq \lambda.
\end{equation}
It can be solved by the same block-wise optimization method, as in the general case. The optimal primal solution is then
$\tilde{\beta}_h=y-R^T\mu$.

\section{Numerical experiments}
\label{s:numerical}

In this section, we present results on a number of simulations and real data studies involving a variety of non-differentiable penalty functions. We compare the alternating linearization algorithm (ALIN) with competing approaches in terms of iteration steps, computation time, and estimation accuracy. All these studies are performed on an AMD 2.6GHZ, 4GB RAM computer using MATLAB.

%\textbf{In this section, we present } In the following experiments we focus on the generalized lasso penalties. This class of penalty functions have been applied to solve many practical problems. In image analysis, total variation based regularization techniques have been widely used for image denoising and deblurring \citep{Beck09, as well as neuroimaging Zhang et al. \cite{Zhang12} to reconstruct copy number variant regions

%This class of penalty functions, including fused lasso and higher order fused lasso penalties, have been widely used for solving problems in image denoising and deblurring \citep{Beck09}, medical imaging , biology \citep{Zhang12}, and dynamic network analysis . In the following, we will demonstrate the effectiveness of the ALIN algorithm for solving this class of problems, particularly for the higher order fused lasso penalty. We compare the alternating linearization algorithm (ALIN) with competing approaches in terms of iteration steps, computation time and estimation accuracy. All these studies are performed on an AMD 2.6GHZ, 4GB RAM computer using MATLAB.

%\lin There have been  generalized lasso penalties have been shown effective in solving The utility of these penalty functions have been demonstrated in

\subsection{$\ell_1$ regularization}

In this section, we compare ALIN with some competing methods for solving the $\ell_1$ regularization problem:
\begin{equation}
\min_\beta \frac{1}{2} \|y-X\beta\|_2^2 + \lambda \|\beta\|_1, \quad \lambda > 0.
\end{equation}
The methods that we are comparing with are: SpaRSA, a type of iterative thresholding method, considered the best method among its variations; FISTA, a variation of the Nesterov method, considered to be state-of-the-art among the first order methods; and SPG, a spectral gradient method. We follow the procedure described by \citep{Wright09} to generate a data set for comparisons.  The elements of the matrix $X$ are generated independently using a Gaussian distribution with mean zero and variance $10^{-2}$. The dimension of $X$ is $n=2^{10}$ by $p=2^{12}$, $p=2^{13}$, and $p=2^{14}$. The true signal, $\beta$, is a vector with $160$ randomly placed $\pm{1}$  spikes and zeros elsewhere. The dependent variables are $y=X\beta + \epsilon$, where $\epsilon$ is Gaussian noise with variance $10^{-4}$.

To make a fair comparison between the methods, we run FISTA on each instance of the problem. FISTA is set to run to $\text{``tol''}=10^{-5}$ or $5,000$ iterations, whichever comes first. Then ALIN, SpaRSA, and SPG are set to run until the objective function values obtained are as good as that of FISTA. We set a parameter $\tau=0.1\|X^Ty\|_\infty$ and  chose values of $\lambda$ =$\tau, 10^{-1}\tau$, $5 \times 10^{-2} \tau$, $10^{-2}\tau$, and $10^{-3}\tau$. We  allow SpaRSA to run its \emph{monotone} and \emph{continuation} feature. Continuation is a special feature of SpaRSA for cases when the parameter $\lambda$ is small. With this feature, SpaRSA computes the solutions for bigger values of $\lambda$ and uses them to find solutions for smaller values of $\lambda$. We did not let SpaRSA use its special feature \emph{de-bias} since it involves removing zero coefficients to reduce the size of the data set. This feature makes it unfair for the other competing methods. In Table \ref{tab:runtime_lasso}, we report the average time elapsed (in seconds) and the standard deviation after $20$ runs.

\begin{table}[h]
\centering
\caption{Average run time (in CPU seconds) and standard deviation (in parenthesis) comparison for combinations of dimension $p$ and tuning parameter $\lambda$. }
\label{tab:runtime_lasso}
\vspace{1ex}
\begin{tabular}[!h]{ccrrrrrr}

\hline
\hline
\hspace{20 mm}&\hspace{20 mm}&\multicolumn{2}{r}{$p=2^{12}$} & \multicolumn{2}{r}{$p=2^{13}$} & \multicolumn{2}{r}{$p=2^{14}$}\\
%\hspace{20 mm}&\hspace{20 mm}&\textbf{CPU (s)}&\textbf{CPU (s)}&\textbf{CPU (s)}&\textbf{CPU (s)}&\textbf{CPU (s)}\\
\hline
$\lambda=\tau$ & \emph{ALIN} & 17.99 & (10.68) & 36.60  & (15.08) &105.14 & (40.88)\\
\hspace{20 mm}& \emph{FISTA} & 8.58 & (4.00) & 18.63 & (9.35) &58.73 & (42.01)\\
\hspace{20 mm}& \emph{SPARSA} & 8.18& (2.34) & 8.18 & (3.80) &  34.23 & (49.55)\\
\hspace{20 mm}& \emph{SPG} & 160.72 & (27.48) &160.72 & (47.82) &404.59 & (79.62)\\

 \cline{2-8}
$\lambda=10^{-1}\tau$ & \emph{ALIN} & 9.35 &  (2.99) &34.01 & (15.18) & 74.06 & (34.27)\\
\hspace{20 mm}& \emph{FISTA} & 16.91 & (4.57) & 36.43 & (16.66) &  131.91 & (33.94)\\
\hspace{20 mm}& \emph{SPARSA} & 20.55 & (10.08) & 36.81 & (19.29)& 169.88 & (73.01)\\
\hspace{20 mm}& \emph{SPG} & 136.26 & (23.30) & 186.94 & (46.56) & 460.71 & (38.93)\\

 \cline{2-8}
$\lambda=5 \times 10^{-2}\tau$ & \emph{ALIN} & 6.30 & (2.73) & 21.83  & (10.17) & 65.79 & (39.49)\\
\hspace{20 mm}& \emph{FISTA} & 18.88 &(2.95) &48.37 & (15.77) &158.73 & (21.86) \\
\hspace{20 mm}& \emph{SPARSA} & 35.56 & (11.32) &74.66 & (24.49) &  234.75 & (64.67)\\
\hspace{20 mm}& \emph{SPG} & 140.00 & (23.15) &190.43 & (45.48) &473.78 & (6.41)\\

 \cline{2-8}
$\lambda=10^{-2}\tau$ & \emph{ALIN} & 4.58 &(2.05) & 16.96 & (10.99) & 28.88 & (16.03)\\
\hspace{20 mm}& \emph{FISTA} & 18.85 & (3.04) & 46.58 & (14.91) & 169.94 & (19.76)\\
\hspace{20 mm}& \emph{SPARSA} & 33.52 & (16.10) & 76.69 &(28.18) & 214.85 & (124.56)\\
\hspace{20 mm}& \emph{SPG} & 140.63 & (22.43) & 196.54 & (43.96) & 483.71 & (4.51)\\

 \cline{2-8}
$\lambda=10^{-3}\tau$ & \emph{ALIN} & 3.68 & (1.20) & 6.67 & (2.43) & 20.16 & (4.31)\\
\hspace{20 mm}& \emph{FISTA} & 18.88 & (2.84) & 45.40 & (14.28) &  162.85 & (36.76)\\
\hspace{20 mm}& \emph{SPARSA} & 19.94 & (12.10) & 39.55 & (27.45)&  92.76 & (102.53)\\
\hspace{20 mm}& \emph{SPG} & 138.73 & (19.56) & 201.74 & (48.09) &467.91 & (101.32) \\

\hline
\hline
\end{tabular}
\end{table}

We can see that the performance of ALIN is comparable to the other methods. In terms of running time, ALIN does better than all competing methods for the range of middle and small values of $\lambda$. For large values of $\lambda$, ALIN does worse than FISTA and SPARSA. For large values of $\lambda$, the solution is fairly close to the starting point $0$, therefore the overhead cost of the update steps and the $f$-subproblem would slow down ALIN. When the value of $\lambda$ reduces, the benefits of these steps become more evident, when ALIN outperforms other methods in terms of running time, by factors of two to three. We should also note that the implementation of FISTA was in $C$, and SpaRSA is a very efficient method specially designed for separable regularization. From our numerical studies, for medium and small values of $\lambda$, FISTA makes very small improvement over $5,000$ iterations and SPARSA has to go through many previous values of $\lambda$ to reach the desired level of objective function values.

\subsection{Fused Lasso regularization}
\label{s:simulation}

In this experiment, we compare the ALIN algorithm with two different approaches using data sets generated from a linear regression model $y=\sum_{j=1}^p x_j \beta_j + \epsilon$, with pre-specified coefficients $\beta_j$, and varying dimen\-sion~$p$.
The values of $x_j$ are drawn from the normal distribution with zero mean and unit variance. The noise  $\epsilon$ is  generated from the normal distribution with zero mean and variance equal to 0.01. Among the coefficients $\beta_j$, 10\% equal 1, 20\% equal 2, and the rest are zero. For instance, with $p=100$, we may have
\[
\beta_j=
        \begin{cases}
          1 & \textup{for}\ j=11,12,\dots, 20, \\
          2 & \textup{for}\ j=21,\dots, 40, \\
          0 & \text{otherwise.}
        \end{cases}
\]

The regularization problem we attempt to solve is
\begin{equation}
\min_\beta \frac{1}{2} \|y-X\beta\|_2^2 + \lambda \sum_{j=2}^p |\beta_j -\beta_{j-1}|, \quad \lambda > 0.
\end{equation}
\begin{table}[h]
\centering
\caption{Run time (in CPU seconds) comparison for combinations of dimension $p$ and tuning parameter $\lambda$ }
\label{tab:runtime1}
\vspace{0.3in}
\begin{tabular}[!h]{ccrrrrr}

\hline
\hline
\hspace{20 mm}&\hspace{20 mm}&{$p=1000$}& {$p=5000$}& {$p$=10,000}& {$p$=20,000}& {$p$=50,000}\\
%\hspace{20 mm}&\hspace{20 mm}&\textbf{CPU (s)}&\textbf{CPU (s)}&\textbf{CPU (s)}&\textbf{CPU (s)}&\textbf{CPU (s)}\\
\hline
$\lambda=10^{-4}$ & \emph{SQOPT} & 10 & 1076 & NA & NA & NA\\
\hspace{20 mm}& \emph{SLEP} & 3 & 1 & 2 & 3 & 5\\
\hspace{20 mm}& \emph{ALIN} & 6 & 0.5 & 1 & 2& 5\\
\hspace{20 mm}& \emph{BREGMAN} & 111 & 24 & 25 & 38& 52\\

 \cline{2-7}

$\lambda=10^{-3}$ & \emph{SQOPT} & 9 & 1025 & NA & NA & NA\\
\hspace{20 mm}& \emph{SLEP} & 4 & 99 & 921 &  2661&4150\\
\hspace{20 mm}& \emph{ALIN} & 9 & 31 & 248 & 665& 1278\\
\hspace{20 mm}& \emph{BREGMAN} & 114 & 21 & 23 & 30& 49\\

 \cline{2-7}
$\lambda=10^{-2}$ & \emph{SQOPT} & 11 & 1019 & NA & NA & NA\\
\hspace{20 mm}& \emph{SLEP} & 2 & 109 & 400 &  1571&6441\\
\hspace{20 mm}& \emph{ALIN} & 6 & 17 & 77 & 313& 815\\
\hspace{20 mm}& \emph{BREGMAN} & 114 & 23 & 57 & 96& 106\\

 \cline{2-7}
$\lambda=0.1$ & \emph{SQOPT} & 11 & 956 & NA & NA & NA\\
\hspace{20 mm}& \emph{SLEP} & 0.4 & 42 & 145 &  508&1358\\
\hspace{20 mm}& \emph{ALIN} & 4 & 22 & 80 & 280& 387\\
\hspace{20 mm}& \emph{BREGMAN} & 103 & 471 & 879 & 2133& 3633\\

 \cline{2-7}
$\lambda=0.2$ & \emph{SQOPT} & 11 & 1015 & NA & NA & NA\\
\hspace{20 mm}& \emph{SLEP} & 1 & 47 & 121 & 387&2251\\
\hspace{20 mm}& \emph{ALIN} & 4 & 52 & 100 & 284& 1360\\
\hspace{20 mm}& \emph{BREGMAN} & 87 & 492 & 833 & 1543& 3541\\

 \cline{2-7}
$\lambda=0.5$ & \emph{SQOPT} & 11 & 1029 & NA & NA & NA\\
\hspace{20 mm}& \emph{SLEP} & 0.9 & 36 & 111 & 386&1584\\
\hspace{20 mm}& \emph{ALIN} & 3 & 47 & 144 & 371& 1730\\
\hspace{20 mm}& \emph{BREGMAN} & 60 & 503 & 924 & 1633& 3543\\

\hline
\hline
\end{tabular}
\end{table}

\begin{figure}[h!]
\centering{
\includegraphics[width=\textwidth]{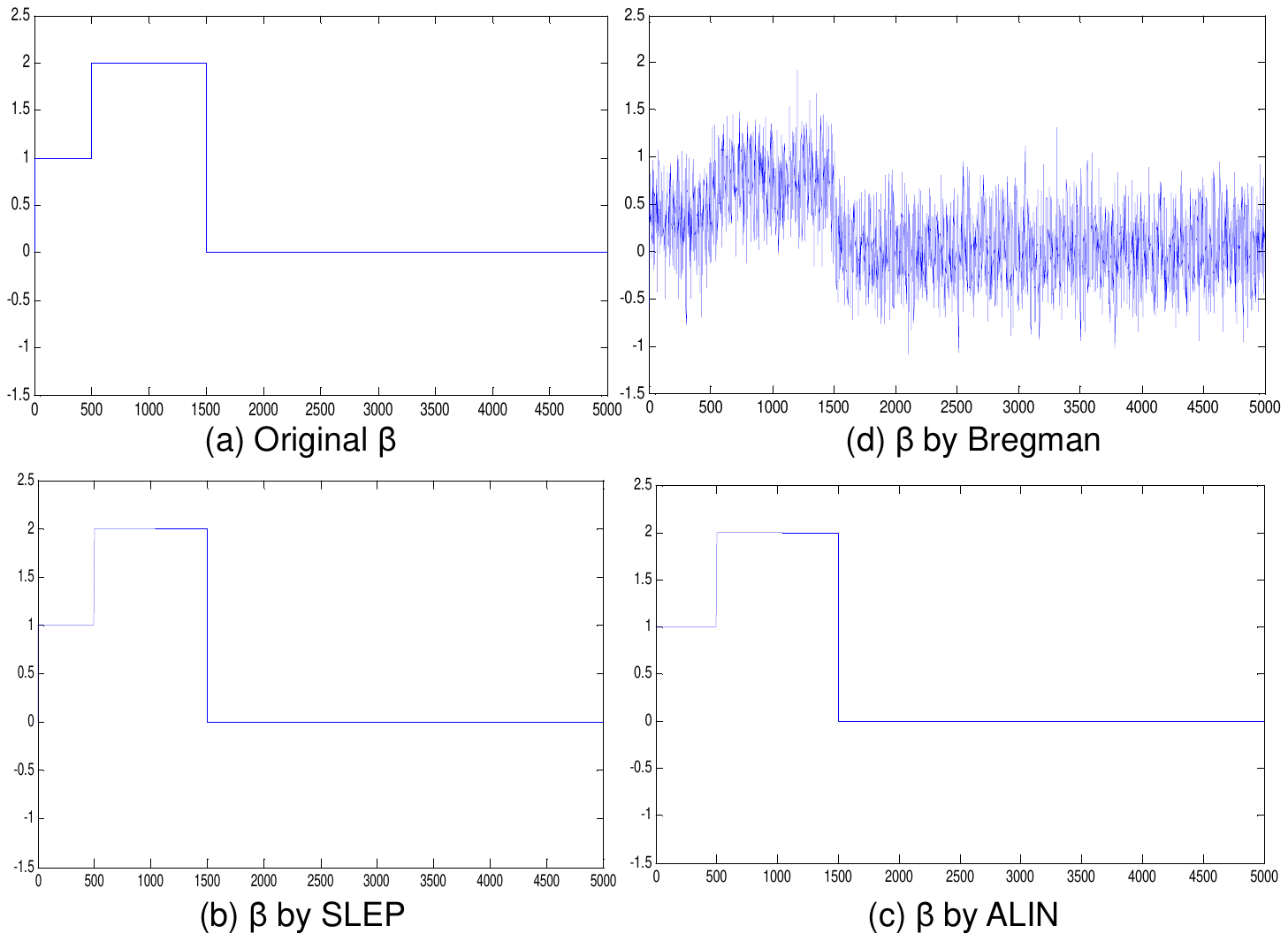}}
\caption{Results of using fused lasso penalty on a simulated data set with $n=1000$, $p=5000$, and $\lambda=0.1$. Plots (a), (b), (c), and (d) correspond to the original $\beta$, results from BREGMAN, SLEP, and ALIN, respectively.}
\label{fig:Simulation}
\end{figure}

Table \ref{tab:runtime1} reports the run times of ALIN and three competing algorithms:  the generic quadratic programming solver (SQOPT), an implementation of Nesterov's method, SLEP, of \citep{Jun11,Nesterov}, and the split Bregman method of \citep{Ye-Xie} (BREGMAN). We fix the sample size at $n$=1000 and vary the dimension $p$ of the problem from 1000 to 50000. Each method is repeated 10 times for different values of the tuning parameter $\lambda$, and the average running time is reported. SLEP's stopping parameter ``tol'' was set to $10^{-5}$. BREGMAN also uses stopping parameter ``tol''= $10^{-5}$. We stop ALIN runs when the objective function value attained is as good as the last value attained by SLEP. Judging from these results, ALIN clearly outperforms the other methods in terms of speed for most cases. The relative improvements on run time can be as much as  8 folds, depending on the experimental setting, and become more significant, when the data dimension grows. This is particularly significant in view of the fact that ALIN was implemented as a MATLAB code, as opposed to the executables in the other cases. Figure \ref{fig:Simulation} presents the solutions obtained by ALIN and SLEP compared to the known parameters. Both ALIN and SLEP achieve results that are very close to the original $\beta$, and the  objective function values are very similar.  BREGMAN performs well when the number of parameters $p$ is not significantly larger than the number of observations $n$.
When $p \gg n$, although BREGMAN has a very good running time, it tends to terminate early and does not provide accurate results. In Fig \ref{fig:Simulation}, we can see that the solution obtained by BREGMAN is not as good as those of SLEP and ALIN.

\begin{figure}
\vspace{-0.6in}
\centering{
\includegraphics[width=0.7\textwidth]{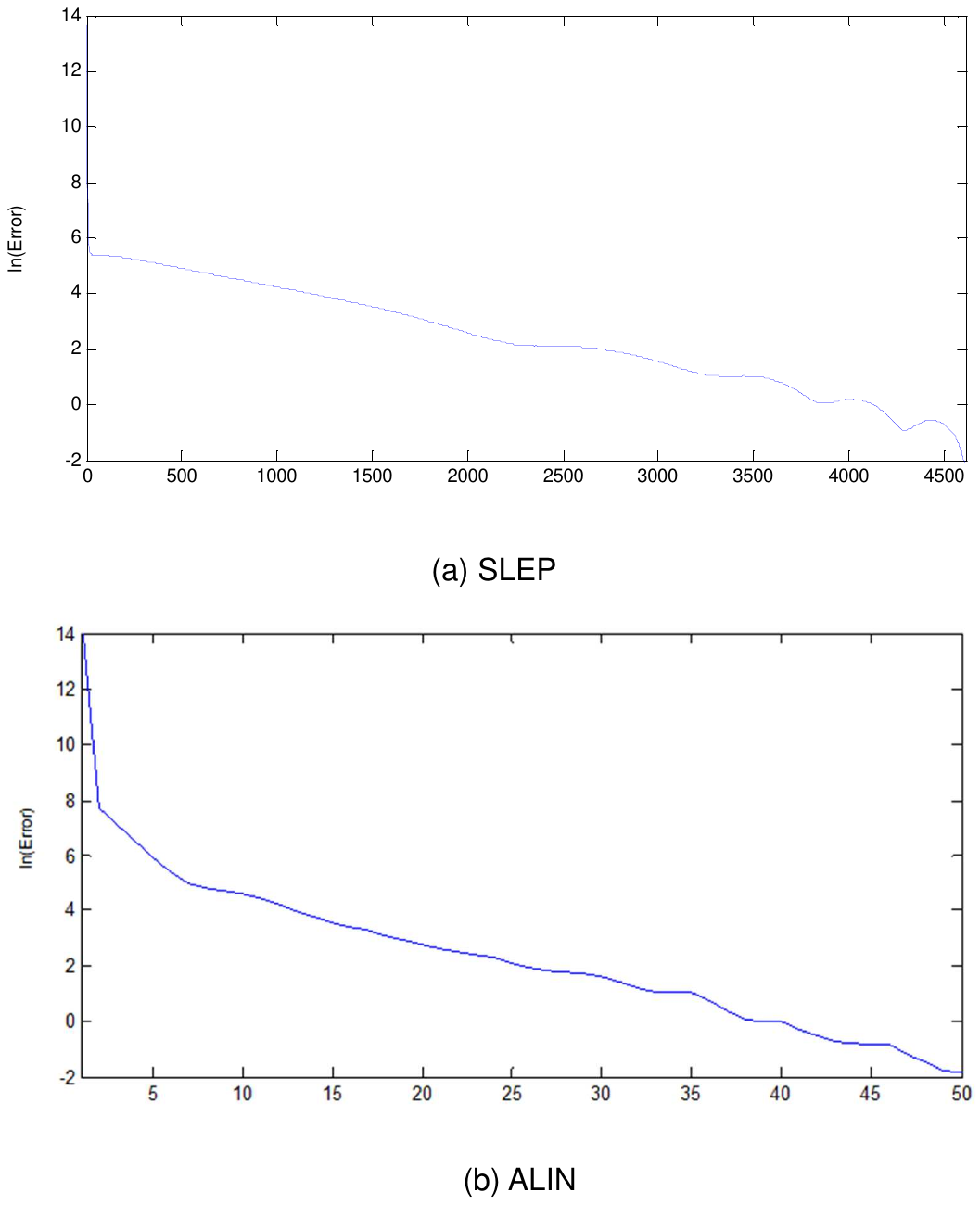}}
\vspace{-0.6in}
\caption{Simulated data set with $n=1000$, $p=5000$, $\lambda=0.1$. Plots (a) and  (b): $\ln(\textup{Error})$ versus iteration number
of SLEP and ALIN, respectively. Error is defined as the difference between the optimal value $\mathcal{L}^{*}$ (obtained by SQOPT) and those obtained by SLEP and ALIN respectively.}
\label{fig:Obj}
\end{figure}

We also investigated how our method approaches the optimal objective function value compared to other methods.
Using the above simulated data set with $n=1000$, $p=5000$, and $\lambda=0.1$, we run ALIN and SLEP to convergence.
At each iteration, we calculated the difference between the optimal value $\mathcal{L}^{*}$ (obtained by SQOPT)
and the current function value for each method.
Figure \ref{fig:Obj} displays (in a logarithmic scale) the progress of both methods. It is clear that ALIN achieves the same accuracy as SLEP in a much smaller number of iterations. Furthermore, the convergence of ALIN is monotonic, whereas that of SLEP is not.

\begin{figure}
\vspace{-0.6in}
\centering{
\includegraphics[width=0.75\textwidth]{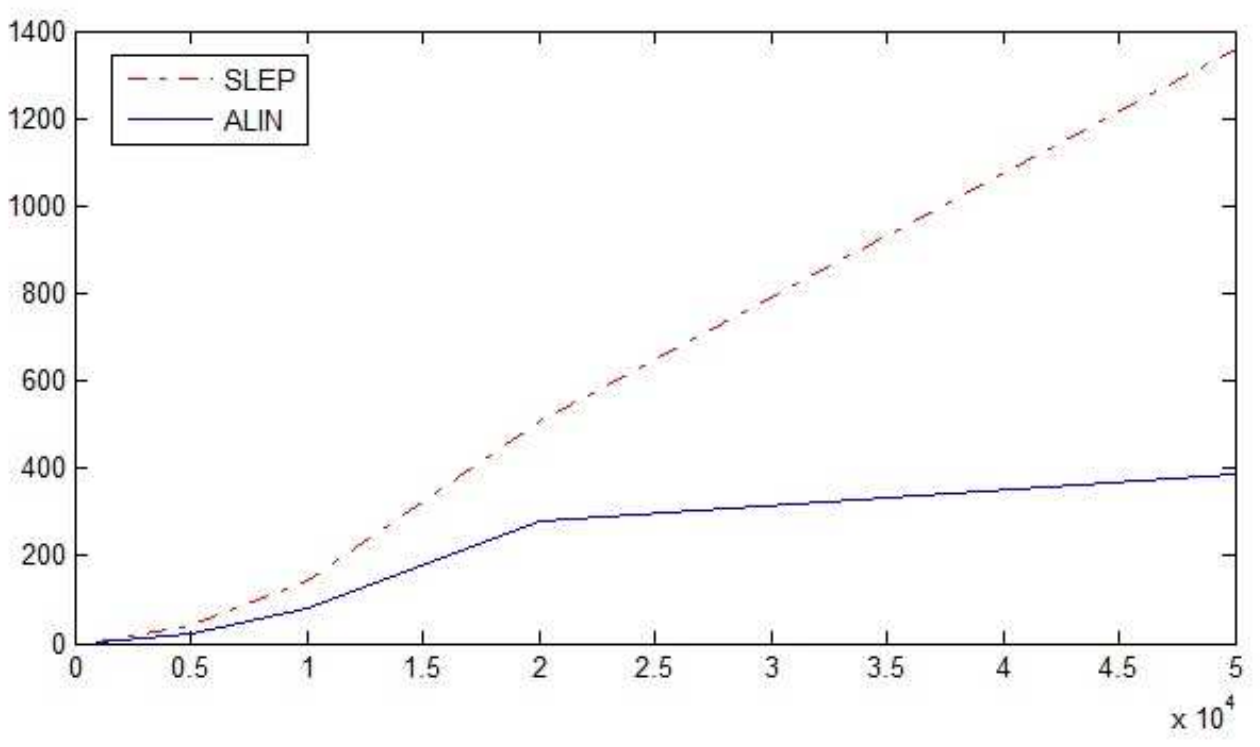}}
\vspace{-1in}
\caption{Running time of SLEP and ALIN as dimension changes. The vertical axis is the run time in seconds, and the horizontal axis is the data dimension.}
\label{fig:Change}
\end{figure}

In Figure \ref{fig:Change} we provide the dependence of the running time of ALIN on the dimensions of the problem,
to illustrate its scalability.
The efficiency of the method is due mainly to its good convergence properties, but also to the efficiency of the preconditioned conjugate gradient method for solving the subproblems. It employs sparse data structures and converges rapidly. Usually, between 10 and 20 iterations
of the conjugate gradient method are sufficient to find the solution of a subproblem.


The update test \eqref{update}  is an essential element of the ALIN method.
For example, in a case with $n=1000$, $p=5000$, and $\lambda=0.1$, the update of $\hat{\beta}$
occurred in about 80$\%$ of the total of 70 iterations, while other iterations consisted only of improving alternating linearizations.
If we allow updates of $\hat{\beta}$ at every step, the algorithm takes more than 5000 iterations to converge in this case.
Similar behavior was observed in all other cases.
These observations clearly demonstrate the difference between the alternating linearization method and the operator splitting methods.

\subsection{CGH data example}

In this study we present the results on analyzing the CGH data using fused lasso penalty. CGH is a technique for measuring DNA copy numbers of selected genes on the genome. The CGH array experiments return the log ratio between the number of DNA copies of the gene in the tumor cells and the number of DNA copies in the reference cells. A value greater than zero indicates a possible gain, while a value less than zero suggests possible losses. \cite{Tibshirani-Wang} applied the fused lasso signal approximator for detecting such copy number variations.
\begin{figure}
\centering
\vspace{-1.in}
\includegraphics[width=\textwidth]{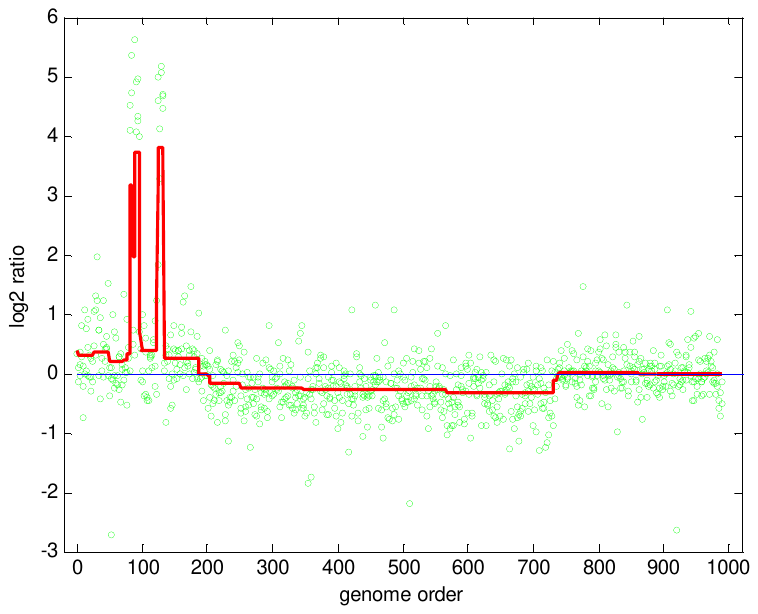}
\vspace{-1.5in}
\caption{Fused lasso applied to CGH data, $\lambda=3$.}\label{fig:cgh}
\end{figure}
This is a simple one-dimensional signal approximation problem with the design matrix $X$ being the identity matrix.
Thus the advantage of ALIN over the other three methods is not significant,
due to the overhead that ALIN has during the conjugate gradient method implemented in MATLAB. Indeed the solution
time of ALIN is comparable to that of Bregman and SLEP.

Figure \ref{fig:cgh}  presents the estimation results obtained by our ALIN method. The green dots  shows the original CNV number, and the red line presents the fused lasso penalized estimates.

\subsection{Total variation based image reconstruction}
\label{s:TV}

In image recovery literature, two classes of regularizers are well known. One is the Tikhonov type of operators, where the regularizing term is quadratic, and the other is the discrete total variation (TV)  regularizer.
The resulting objective function from the first type is relatively easy to minimize, but it  tends to over-smooth the image, thus failing to preserve its sharpness \citep{Wang08}. In the following experiment, we demonstrate the effectiveness of ALIN in solving TV-based image deblurring problems,   with discrete TV, as well as a comparison to the Tikhonov regularizer.

Although of similar form, higher-order fused lasso models are fundamentally different from the one-dimensional fused lasso, as the structural matrix $R$ appearing in eq. \eqref{glasso1} is not full-rank and $R^{T}R$ is ill-conditioned. This additional complication introduces considerable challenges in the path type algorithms \citep{tib11}, and additional computational steps need to be implemented to guarantee convergence. The ALIN algorithm does not suffer from complications due to the singularity of $R$,
because the dual problem \eqref{Rdual} is always well-defined and has a solution.
Even if the solution is not unique, \eqref{fl-f} is still an optimal solution of the $h$-subproblem, and the algorithm proceeds unaffected.

 Let ${y}$ be an $m \times n$ observed noisy image; one attempts to minimize the following objective function:
 \begin{equation}
 \label{TV-model}
\mathcal{L}(\beta)= \half \| {y}-\mathcal{A}(\beta)\|_2^2 + \lambda h({\beta}),
\end{equation}
where $h(\beta)$ is an image variation penalty,
and $\mathcal{A}: {\R}^{m \times n} \rightarrow {\R}^{m \times n}$  is a linear transformation. When $\mathcal{A} $ is the identity transformation, the problem is to \emph{denoise} the image $y$, but we are rather interested in a significantly more challenging problem of
\emph{deblurring}, where $\mathcal{A} $ replaces each pixel with the average of its neighbors and itself
(typically, a 3 by 3 block, except for the border).

The penalty can be defined
as the $\ell_1$-norm of the differences between neighboring pixels (\emph{ $\ell_1$-TV}),
\begin{equation}
\label{TV1}
h({\beta})= \sum_{i=1}^{m-1} \sum_{j=1}^{n-1} \big( |\beta_{i,j}-\beta_{i+1,j}| + |\beta_{i,j}-\beta_{i,j+1}|\big)
+\sum_{i=1}^{m-1} | \beta_{i,n}-\beta_{i+1,n}| + \sum_{j=1}^{n-1} |\beta_{m,j}-\beta_{m,j+1}|,
\end{equation}
or as follows (\emph{$\ell_2$-TV}):
\begin{equation}
\label{TV2}
h({\beta})= \sum_{i=1}^{m-1} \sum_{j=1}^{n-1}
\big( |\beta_{i,j}-\beta_{i+1,j}|^2 + |\beta_{i,j}-\beta_{i,j+1}|^2\big)^{1/2}
+\sum_{i=1}^{m-1} | \beta_{i,n}-\beta_{i+1,n}| + \sum_{j=1}^{n-1} |\beta_{m,j}-\beta_{m,j+1}|.
\end{equation}
It is clear that both cases can be cast into the general form \eqref{glasso1}, with the operator $R$ representing
the evaluation of the differences $\beta_{i,j}-\beta_{i+1,j}$ and $\beta_{i,j}-\beta_{i,j+1}$. The regularizing
function \eqref{TV1} corresponds to the $\ell_1$-norm of $R\beta$, while the function \eqref{TV2}
corresponds to a
norm of form \eqref{norm12}. In the latter case, we have $mn$ blocks, each of dimension two, except for the border blocks,
which are one-dimensional.

In the following experiments, we apply the $\ell_1$-TV to recover noisy and blurred images to their original forms. The resulting regularization problems are rather complex. Deblurring a 256 by 256 image results in solving a very large generalized lasso problem
(the matrix $R$ has dimensions of about $262000 \times 66000$). The $f$-subproblem is solved using the block coordinate descent method and the $h$-subproblem is solved using the conjugate gradient method with $\text{``tol''}=10^{-5}$, as discussed previously. The fact that $A$ and $R$ are sparse matrices makes the implementation very efficient, as demonstrated in the numerical study.

\begin{figure}
\centering
\includegraphics[width=1.2\textwidth]{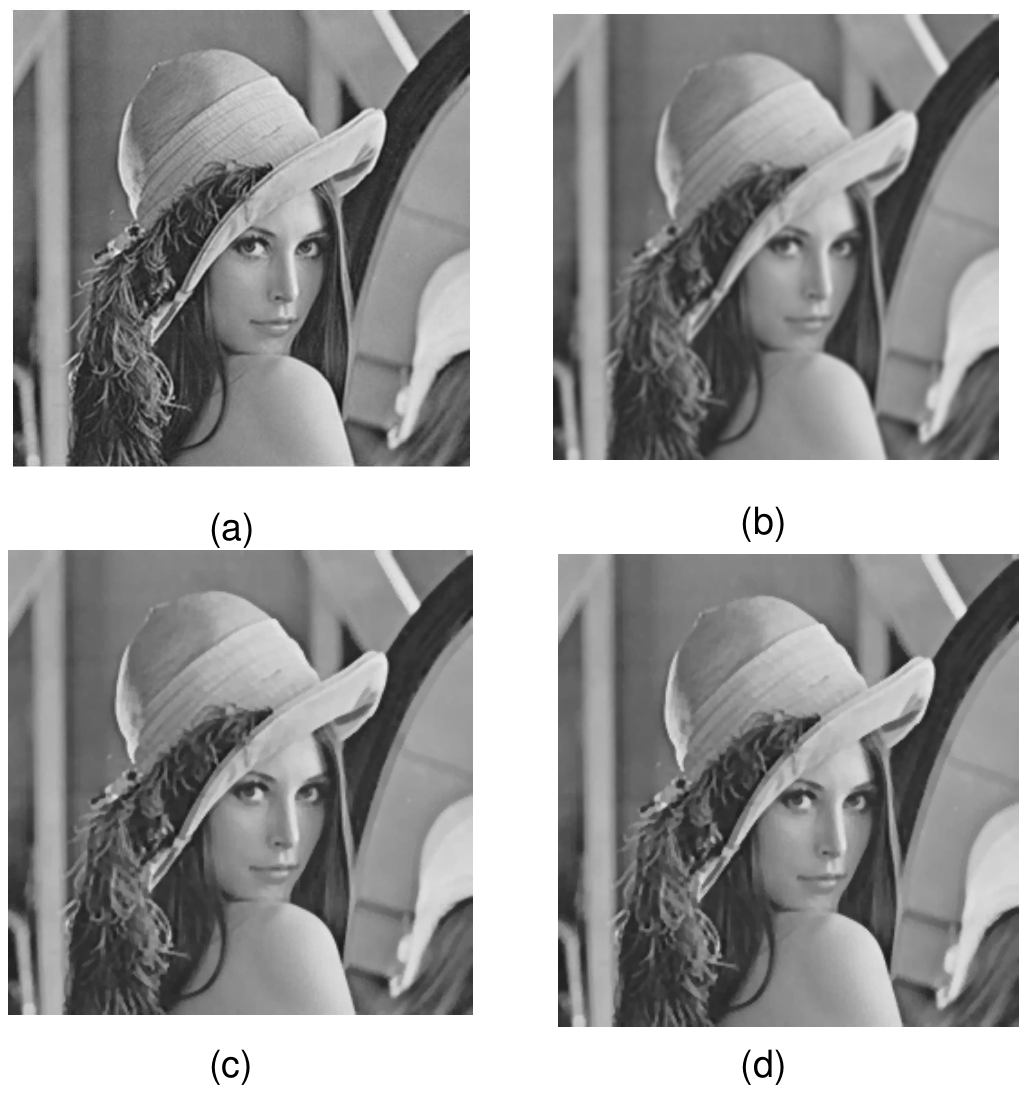}
\vspace{-0.4in}
\caption{Results of deblurring using fused lasso penalty. Plots (a), (b), (c), and (d)  correspond to the original image, the blurred image,  the ALIN de-blurred image, and the FISTA de-blurred image, respectively.}\label{fig:deblurr}
\end{figure}

First, we \emph{blur} the image,
by replacing each pixel with the average of its neighbors and itself.
This operation defines the kernel operator $\mathcal{A}$ used in the loss function
$\half \| {y}-\mathcal{A}(\beta)\|_2^2$.
Then we add $N(0,0.02)$ noise to each pixel. Clearly, for image deblurring, the design matrix is no longer the identity matrix, thus the problem
 is more complicated than the image denoising problem. The deblurring results on a standard example (``Lena'')
are shown in Figure \ref{fig:deblurr}; similar deblurring results from ALIN and FISTA are observed.

\begin{figure}[h]
\begin{center}$
\begin{array}{cc}
\hspace{-2.5em}\includegraphics[width=3.5in]{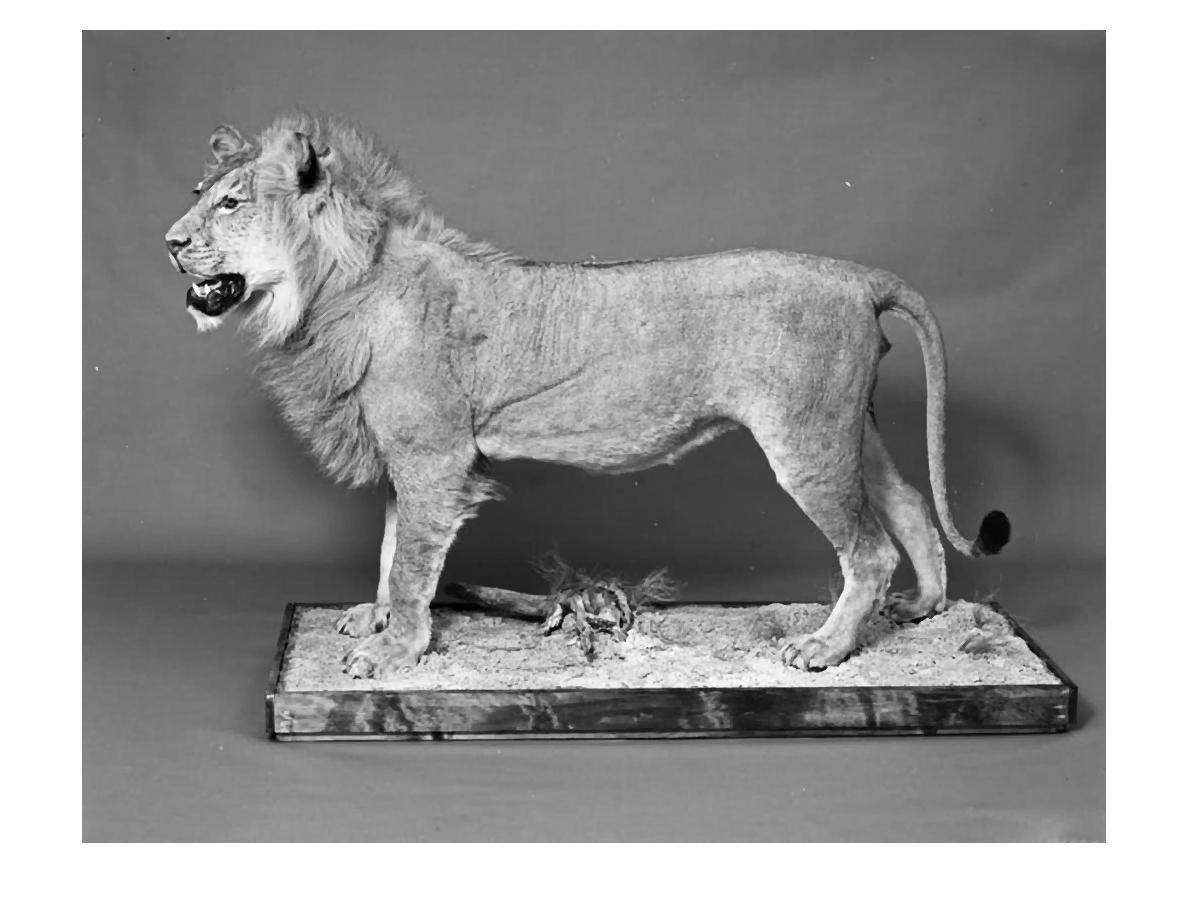} &
\includegraphics[width=3.5in]{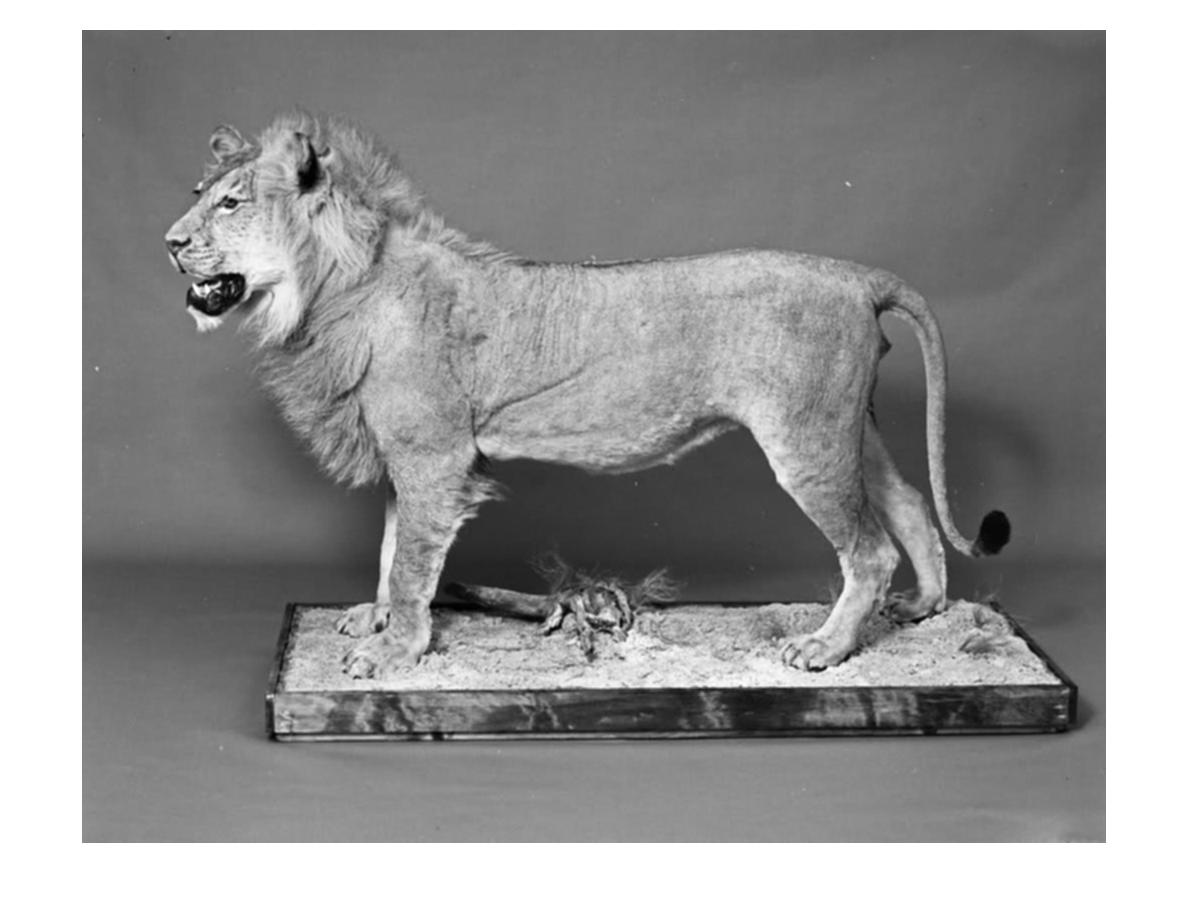}
\end{array}$
\end{center}
\caption{Image deblurring on the ``lion'' data. The left plot is the result from the $\ell_1$-TV penalty; the right plot is from the Tikhonov penalty.} \label{fig:deblurr2}
\end{figure}

Next, we run the image deblurring on a 1 Megapixel image. We compare the result of image deblurring using the $\ell_1$-TV and a quadratic Tiknonov regularization approach, which corresponds to
formula \eqref{TV2} without the square root operations:
\begin{equation}
\label{TV3}
h({\beta})= \sum_{i=1}^{m-1} \sum_{j=1}^{n-1}
 |\beta_{i,j}-\beta_{i+1,j}|^2 + |\beta_{i,j}-\beta_{i,j+1}|^2
+\sum_{i=1}^{m-1} | \beta_{i,n}-\beta_{i+1,n}|^2 + \sum_{j=1}^{n-1} |\beta_{m,j}-\beta_{m,j+1}|^2.
\end{equation}
 The results are shown in
 Figure~\ref{fig:deblurr2}. It is seen that the $\ell_1$-TV recovers a sharper image than the quadratic penalty. Deblurring with the two-dimensional fused lasso penalty yields an MSE of 7.2 with respect to the original image, while that of Tikhonov regularization is 9.3. Deblurring with the regularizer \eqref{TV1} has an almost identical effect as with \eqref{TV2}.

There have been many efficient iterative methods proposed to solve this problem. Two outstanding general frameworks are a variation of Nesterov's gradient method \citep{Nesterov} and the method of alternating direction (ADMM). SLEP is a variation of Nesterov method like FISTA, although it was specifically implemented for fused-lasso penalty. It is not directly applicable for total variation deblurring problem. We pick two algorithms to compare with ALIN in this numerical study: FISTA of \cite{Beck09}, a very efficient first-order method for discrete total variation based image processing; and TVAL, a method based on Augmented Lagrangian and Alternating  Direction algorithm.
TVAL solves a model equivalent to \eqref{TV-model}, but with a coefficient $\mu$ in front of the least-squares term, instead of $\lambda$
 at the regularization. %different model:
%\begin{equation}
%\min_u \sum_i \|D_i u\|_1 + \frac{\mu}{2}\|Au -b\|_2 ,
%\end{equation}
%where $b$ is the blurred and noisy observed image, $A$ is a linear blurring operator as defined above, and $D$ is the total variation matrix.
%However, by picking $\mu=\frac{1}{\lambda}$, we obtain the same model.

In the first comparison, we pick $10$ random grayscale images with small size, typically $205 \times 205$, or approximately $40,000$ pixels. Following the same procedure as described in \citep{Beck09} the image is blurred using a $3 \times 3$ kernel and a Gaussian noise with variance $10^{-2}$ is added. The deblurring procedure is run with a few different values for $\lambda$. We let FISTA runs $100$ iterations with the $monotone$ feature, which keeps the objective function decrease monotonically, and the tolerance is set to $10^{-5}$. Then we run ALIN to the same objective function value. For TVAL, unfortunately, we cannot proceed similarly. Thus we let TVAL run $10,000$ iterations or to $\text{``tol''}=10^{-5}$, whichever comes first. To compare the quality of the restored image, we use the signal-to-noise (SNR) ratio defined as

\begin{equation}
SNR=10\log10\frac{\|u^0-\tilde{u}\|^2}{\|u^0-u\|^2},
\end{equation}
where $u^0$ is the original image, $\tilde{u}$ is the mean intensity of the original image, and $u$ is the restored image. In Table \ref{t:3}, we report the running time to produce the best quality restored image, where the regularization parameter $\lambda=10^{-4}$, similar to what was suggested by \citep{Li13}. We also report SNR and the mean squared error (MSE).

\begin{table}[htbp]
\centering
\caption{Run time comparison on image deblurring - small size images.}
\vspace{0.1in}
\begin{tabular}{|c|c|c|c|}
\hline
Method & CPU time (secs) & SNR & MSE\\
\hline
FISTA & 9.19 & 11.00 & 4.21 \\
ALIN & 6.85 & 11.03 & 4.18 \\
TVAL & 3.03 & 10.56 & 4.57 \\
\hline
\end{tabular}
\label{t:3}
\end{table}

\begin{figure}[htbp]
\centering
\includegraphics[width=0.75\textwidth]{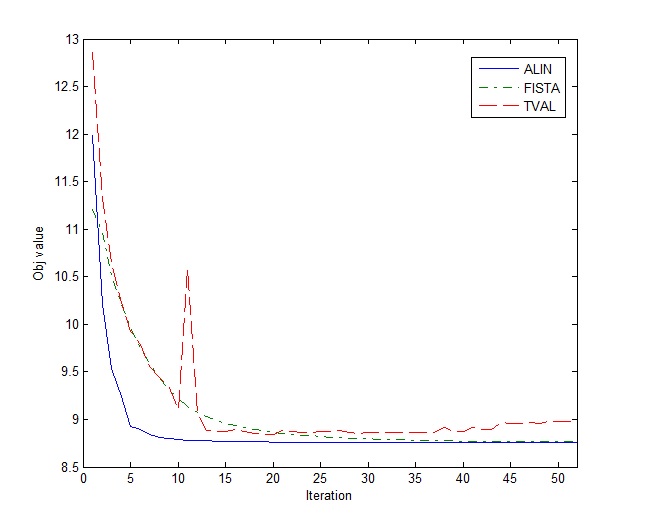}
\caption{Progression in terms of objective function values of ALIN, FISTA, and TVAL}\label{fig:obj_image}
\end{figure}

Although TVAL has superior performance in terms of running time, when compared to FISTA and ALIN, it produces an image of lower quality. With the same value of parameter $\lambda$, TVAL was not able to obtain the same objective function value as FISTA and ALIN. This makes the SNR of the TVAL-restored image lower and the error higher than those of ALIN and FISTA. ALIN and FISTA have similar performance in terms of image quality, but ALIN is more efficient than FISTA.
In Figure~\ref{fig:obj_image}, we plot the progression in terms of objective function values for all three methods. TVAL takes only 52 iterations to terminate. In this plot, ALIN and FISTA are set to terminate in 52 iterations.

In the second comparison, we pick $10$ random grayscale images with medium size, ranging from $200,000$ to $500,000$ pixels. The experiment is carried out in the same manner as the previous one. The results are reported in Table \ref{t:31}, and we observe the same pattern as in the previous comparison.

\begin{table}[htbp]
\centering
\caption{Run time comparison on image deblurring - medium size images.}
\vspace{0.1in}
\begin{tabular}{|c|c|c|c|}
\hline
Method & CPU time (secs) & SNR & MSE\\
\hline
FISTA & 65.41 & 12.14 & 5.98 \\
ALIN & 41.28 & 12.14 & 5.97 \\
TVAL & 7.18 & 8.30 & 14.36 \\
\hline
\end{tabular}
\label{t:31}
\end{table}

\subsection{Application to a narrative comprehension study for children}
\label{s:children}

With high dimensional fused lasso penalty, the constrained optimization problem with identity design matrix is already difficult to solve,
and a large body of literature has been devoted to solving this problem. When the design matrix is not full rank, the problem becomes much more difficult. In this section, we apply the three-dimensional fused lasso penalty to an regression problem where the design matrix $X$ contains many more columns than rows.

%However, there has not been a  systematic treatment of penalized regression using non-separable penalty functions with an arbitrary design matrix $X$ that may have non-full rank.

Specifically, we perform regularized regression between the measurement of children's language ability (the response $y$) and voxel level brain activity during a narrative comprehension task (the design matrix $X$). Children develop a variety of skills and strategies for narrative comprehension during early childhood years \citep{prasa10}. This is a complex brain function that involves various cognitive processes in multiple brain regions. We are not attempting to solve the challenging neurological problem of identifying all such brain regions for this cognitive task.
Instead, the goal of this study is to demonstrate ALIN's ability for solving constrained optimization problems of this type and magnitude.

The functional MRI data are collected from 313 children with ages 5 to 18 \citep{Schmithorst06}. The experimental paradigm is a 30-second block design with alternating stimulus and control. Children are listening to a story read by adult female speaker in each stimulus period, and pure tones of 1-second duration in each resting period. The subjects are instructed to answer ten story-related multiple-choice questions upon the completion of the MRI scan (two questions per story). The fMRI data were preprocessed and transformed into the Talairach stereotaxic space by linear affine transformation. A uniform mask is applied to all the subjects so that they have measurements on the same set of voxels.

The response variable $y$ is the oral and written language scale (OWLS). The matrix $X$ records the activity level for all the 8000 voxels measured. The objective function is the following:
\[
\mathcal{L}(\beta) = \half \| y- X \beta \|^2_2 +\lambda_1 h_1(\beta) +\lambda_2 h_2(\beta),
\]
where
\begin{align*}
h_1({\beta})&=\sum_{i=1}^{m-1} \sum_{j=1}^{n-1}\sum_{k=1}^{p-1} \{ |\beta_{i,j,k}-\beta_{i+1,j,k}| +
|\beta_{i,j,k}-\beta_{i,j+1,k}|+|\beta_{i,j,k}-\beta_{i,j,k+1}| \} \\
&{\quad} + \sum_{i=1}^{m-1} \sum_{k=1}^{p-1} \{ | \beta_{i,n,k}-\beta_{i+1,n,k}| +| \beta_{i,n,k}-\beta_{i,n,k+1}| \} + \sum_{j=1}^{n-1}\{
|\beta_{m,j,p}-\beta_{m,j+1,p}| \} \\
&{\quad} +\sum_{j=1}^{n-1}\sum_{k=1}^{p-1} \{| \beta_{m,j,k}-\beta_{n,j+1,k}| +| \beta_{m,j,k}-\beta_{m,j,k+1}| \}+\sum_{i=1}^{m-1}\{
|\beta_{i,n,p}-\beta_{i+1,n,p}| \} \\
&{\quad} +\sum_{i=1}^{m-1}\sum_{j=1}^{n-1} \{| \beta_{i,j,p}-\beta_{i+1,j,p}| +| \beta_{i,j,p}-\beta_{i,j+1,p}| \} + \sum_{k=1}^{p-1}\{
|\beta_{m,n,k}-\beta_{m,n,k+1}| \},\\
h_2(\beta) &= \sum_{i=1}^{m} \sum_{j=1}^{n}\sum_{k=1}^{p}  |\beta_{i,j,k}|,
\end{align*}
and $m=31$, $n=35$, and $p=15$.

%Clearly, the design matrix does not have full rank as the dimension is far greater than the sample size. None of the methods that we compared previously work in  this setting with a 3-d fused lasso penalty. The ALIN algorithm terminates in 149 steps and the run time is 575 seconds.

\begin{figure}[h]
\vspace{-0.6in}
\centering
\includegraphics[trim=0cm 14cm 0cm 0cm,width=\textwidth]{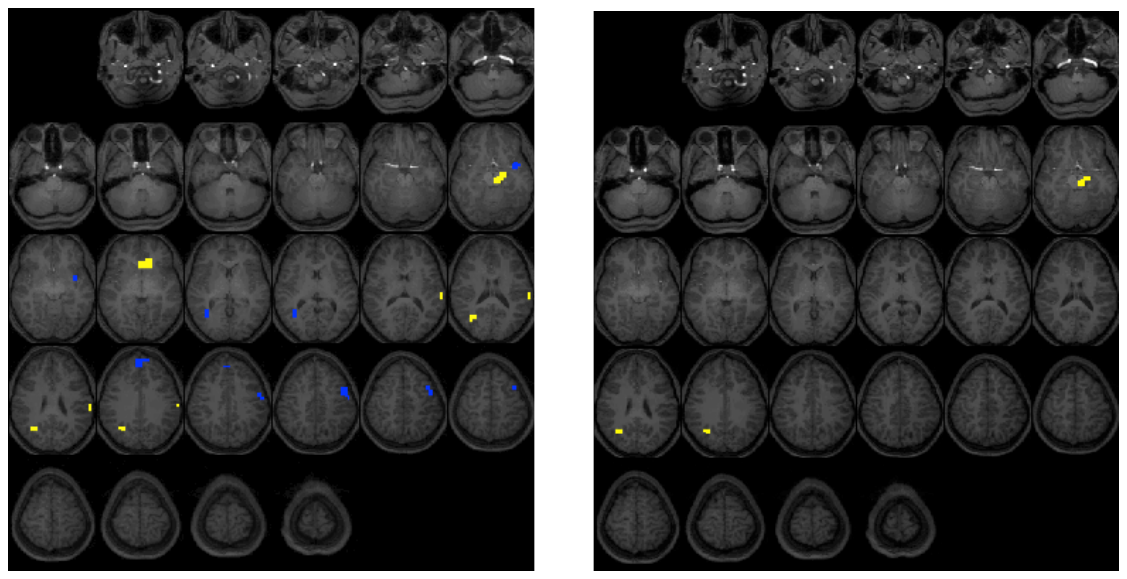}
\vspace{-0.6in}
\caption{ Results of regularization regression with combined lasso and 3-d fused lasso penalty. The tuning parameters of fused lasso is 0.2 for both figures. The tuning parameter for lasso is 0.2 for the left and 0.6 for the right.}\label{fig:fuse1}
\end{figure}

While the main purpose of this study is to demonstrate the capability of the ALIN algorithm for solving penalized regression problems with 3-d fused lasso, there are also some interesting neurological observations. One objective of this study is to identify the voxels that are significant for explaining the performance score $y$. These voxels constitute active brain regions that are closely related to the OWLS. Figure \ref{fig:fuse1} presents the results of fitted coefficients using combined lasso and fused lasso penalty. The highlighted regions shown in the maps are areas with more than 10 voxels (representing clusters of size 10 and above). The left plot in the figure is the optimal solution obtained using ten-fold cross validation. The optimal tuning parameters are 0.2 for both fused lasso and lasso penalties. Roughly speaking, five brain regions have been identified. The yellow area to the rightmost side of the brain is situated in the wernicke area, which is one of the two parts of the cerebral cortex linked to speech. It is involved in the understanding of written and spoken language. The only difference between the left and right plots is the value of the tuning parameter for the lasso penalty, which is 0.2 and 0.6 respectively.
Clearly, the right plot shrinks more coefficients to zero, which results in a reduced number of significant regions, as  compared to the left plot.

 \begin{figure}[h]
 \vspace{-0.6in}
\centering
\includegraphics[trim=0cm 14cm 0cm 0cm,width=\textwidth]{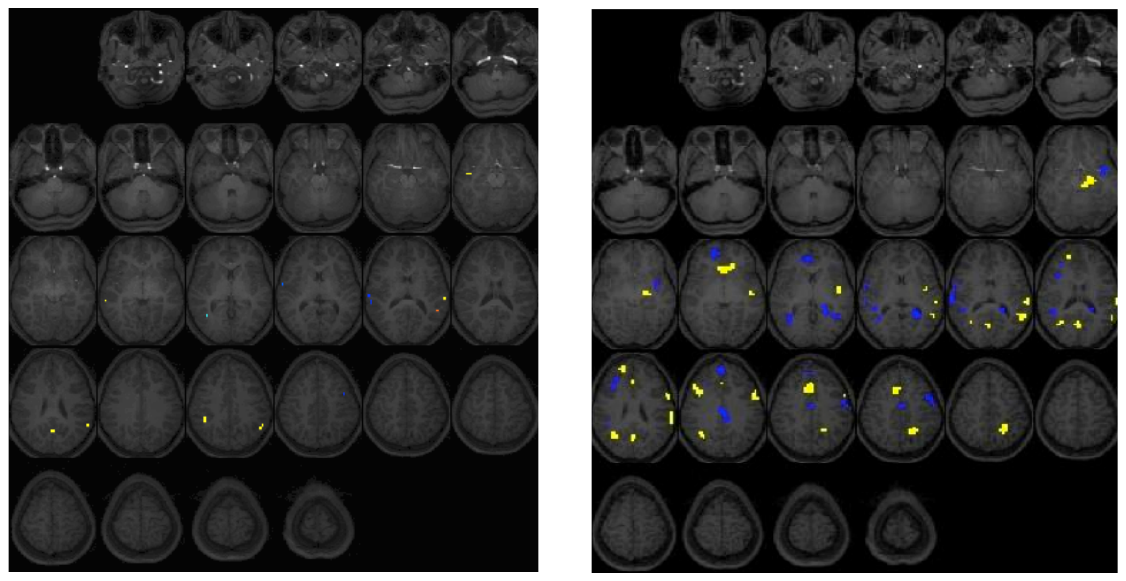}
\vspace{-0.6in}
\caption{ Results of regularization regression with lasso penalty (left plot) and Tiknonov type penalty (right plot). }\label{fig:lassoandl2}
\end{figure}

We further study this regularization problem using only lasso penalty and Tiknonov type penalty similar to \eqref{TV3}.  Figure \ref{fig:lassoandl2} shows the fitted maps. The left plot is the case where only lasso penalty is applied. Comparing this with Figure \ref{fig:fuse1}, we see that the 3-d fused lasso penalty imposes smoothing constraints on the neighboring coefficients, thus allowing to identify larger areas significant for the response variable $y$. The simple lasso penalty imposes shrinkage on the coefficients individually, resulting in rather disjoint significant voxels. Such scatterness is much less informative for neurologists than larger areas identified by the three-dimensional fused lasso penalty. Meanwhile, the Tiknonov type regularization generates too many significant regions as shown in the right plot. This is partially due to the over smoothing of the image as discussed in the previous section.

For comparison, we have considered a couple of methods designed to solve this particular problem. \emph{Genlasso} is the path algorithm designed for the Generalized Lasso in the original paper. However, it was unable to handle an instance of this magnitude.

Another method is the Augmented Lagrangian and Alternating Direction method. The problem of interest
\begin{equation}
\min_\beta \frac{1}{2} \|y - X\beta\|^2_2 + \lambda\|R\beta\|_1, \quad \lambda > 0
\end{equation}
can be reformulated as
\begin{equation}
\min_\beta \frac{1}{2} \|y - X\beta\|^2_2 + \lambda \|z\|_1 \quad s.t: R \beta -z = 0 .
\end{equation}

The Augmented Lagrangian is:
\begin{equation}
\mathcal{L}(\beta,z,u)= \frac{1}{2} \|y - A\beta\|^2_2 +  \lambda \|z\|_1 + \rho u^T(R\beta -z) + \frac{\rho}{2}\|R\beta -z\|_2^2,
\end{equation}
where $\rho>0$ is the penalty coefficient.
This formulation can be solved by the Alternating Direction method. We implemented this method in Matlab. The update step for $\beta$ requires minimizing a quadratic function. The iterative method of choice to solve the sub-problems is the conjugate gradient method built in Matlab. The performance of this method strongly depends on the choice of the penalty parameter $\rho$. As suggested in \citep{Wahlberg12}, we choose $\rho=\lambda$ to keep the algorithm stable for the implementation.
It is known that the method has a nice decrease in the function values but slow tail convergence and an iteration of ADMM for this particular problem is rather expensive so we let it run for $30$ iterations and let ALIN run until it reached the same objective function value. The running times for different values of $\lambda$ are reported in Table~\ref{t:32}.

\begin{center}
\begin{table}[htbp]
\centering
\caption{Run time comparison on 3D fused lasso.}
\vspace{0.1in}
\begin{tabular}{|c|c|c|c|c|c|c|}
\hline
Method & $\lambda= 0.001$ & $\lambda=0.01$ & $\lambda=0.05$ &  $\lambda=0.1$ & $\lambda=1$\\
\hline
	& CPU time (secs) & CPU time (secs)  & CPU time (secs)  & CPU time (secs)  & CPU time (secs)  \\
ALIN &  20.68 &  12.19 & 18.58 & 23.45 & 129.81 \\
ADMM & 72.86 & 94.69 & 68.13 & 74.43 & 267.01 \\
\hline
\end{tabular}
\label{t:32}
\end{table}
\end{center}

%\begin{center}
%\begin{table}[htbp]
%\centering
%\caption{Run time comparison on 3D fused lasso.}
%\label{deblur}
%\vspace{0.1in}
%\begin{tabular}{|c|c|c|c|c|c|c|}
%\hline
%Method & $\lambda=1e-3$ & $\lambda=5e-3$ & $\lambda=1e-2$ & $\lambda=5e-2$ &  $\lambda=1e-1$ & $\lambda=1$\\
%\hline
%	& CPU time (secs) & CPU time (secs)  & CPU time (secs)  & CPU time (secs)  & CPU time (secs)  & CPU time (secs)  \\
%ALIN &  20.68 & 11.47 & 12.19 & 18.58 & 23.45 & 129.81 \\
%ADMM & 72.86 & 105.46 & 94.69 & 68.13 & 74.43 & 267.01 \\
%\hline
%\end{tabular}
%\label{t:3}
%\end{table}
%\end{center}

We also implemented FISTA for this particular problem. For each iteration $k$ of FISTA, the following optimization problem is solved:
\begin{equation}
\min_\beta
\mathcal{Q}_L(\beta,\beta^k) = f(\beta^k) +  \langle \nabla f(\beta^k),\beta-\beta^k \rangle + \frac{L}{2}\|\beta-\beta^k\|^2 + g(\beta),
\end{equation}
where $f$ is Gaussian loss function and $g(\beta)=\|R\beta\|_1$. The parameter $L$ is the Lipschitz constant of $\nabla f $.
This problem is similar to the f-subproblem of ALIN, so we utilize our own block-coordinate descent method to solve this problem. Since the Lipschitz constant $L$ cannot be computed efficiently, we use back-tracking to find the proper $L$. Back-tracking is a popular method to find the right step size for FISTA iterations, however it can slow down the algorithm significantly. We observe that in each iteration, back-tracking will have to solve the sub-problem $20$ to $30$ times to find a good approximation to the Lipschitz constant of $\nabla f$. Normally, this quantity is approximated by the maximum eigenvalue of $X^TX $. However, in the $p \gg n $ setting, it is not computationally efficient to estimate eigenvalues. When FISTA is applied to this data set with 3-d fused lasso penalty, it needs around $10,000$ iterations to reach the same objective function value as ADMM, and the running time is more or less an hour.  This is also in line with what we observe from the implementation of FISTA for 1-d fused lasso penalty in the package SPAMS.

\section{Conclusion}
\label{s:conclusions}

The alternating linearization method is a specialized nonsmooth optimization method for solving structured nonsmooth optimization problems. It combines the ideas
of bundle methods and operator splitting methods, to define a descent algorithm in terms of the values of the function that is minimized. We have adapted
the alternating linearization method to structured regularization problems by
 introducing the idea of diagonal quadratic approximation and developing specialized methods for solving subproblems.
 As a result, a new general method for a variety
 of regularization problems has been obtained, which has the following theoretical features:
 \begin{itemize}
 \item It deals with nonsmoothness directly, not via approximations,
 \item It is monotonic with respect to the values of the function that is minimized,
 \item Its  convergence is guaranteed theoretically.
 \end{itemize}

 Our numerical experiments on a variety of structured regularization problems illustrate the applicability of the alternating linearization method
 and indicate its practically important virtues: speed, scalability, and accuracy. It clearly outperforms extant methods, and it can solve problems which were unsolvable otherwise.

 Its efficacy and accuracy follow from the use of the diagonal quadratic approximation and from a special test,
 which chooses in an implicit way the best operator splitting step to be performed. The current approximate solution is updated only if it leads to
 a significant decrease of the value of the objective function.

 Its scalability is due to the use of highly specialized algorithms
 for solving its two subproblems.  The algorithms do not require any explicit matrix formation or inversion, but only matrix--vector multiplications,
 and can be efficiently implemented with sparse data structures.

Our study of image denoising and deblurring in section \ref{s:TV} , as well as the narrative comprehension for children in section \ref{s:children} are illustrations of broad applicability of the alternating linearization method.
 %It is, to the best of our knowledge,  the first successful method for this three-dimensional fused lasso problem.

%\section*{Acknowledgements} The authors are grateful to the Associate Editor and two anonymous Referees, whose insightful comments helped to make significant improvements in the results reported in this manuscript.

%\bibliographystyle{plainnat}
\bibliography{AL-AR}

\end{document}